\documentclass[aps,prd,onecolumn,groupedaddress,showpacs,nofootinbib,amssymb]{revtex4}
\usepackage[dvips]{graphicx}
\usepackage{amssymb}
\usepackage{amsmath}
\usepackage{graphicx,,color}
\usepackage{amsfonts}
\usepackage{bm}
\usepackage{cancel}
\usepackage{comment}

\newcommand\be{\begin{equation}}
\newcommand\ee{\end{equation}}

\allowdisplaybreaks[4]

\begin{document}

\tolerance=5000

\title{Generalized Horndeski-Like Einstein Gauss-Bonnet Inflation with Massless Primordial Gravitons}
\author{V.K.~Oikonomou,$^{1,2,3}$\,\thanks{v.k.oikonomou1979@gmail.com}F.P.
Fronimos,$^{1}$\,\thanks{fotisfronimos@gmail.com}} \affiliation{
$^{1)}$ Department of Physics, Aristotle University of
Thessaloniki, Thessaloniki 54124,
Greece\\
$^{2)}$ Laboratory for Theoretical Cosmology, Tomsk State
University of Control Systems and Radioelectronics, 634050 Tomsk,
Russia (TUSUR)\\
$^{3)}$ Tomsk State Pedagogical University, 634061 Tomsk,
Russia\\}

\tolerance=5000

\begin{abstract}
In this work we shall introduce a theoretical framework comprised
by a non-minimal coupled canonical scalar field, a non-minimal
coupling to the Gauss-Bonnet invariant and a non-minimal kinetic
coupling. This theoretical framework is basically a non-minimally
coupled Einstein-Gauss-Bonnet theory with extra corrections of the
form of a non-minimal kinetic coupling. In order to comply with
the GW170817 event, we shall impose a constraint on the
propagation speed of the primordial tensor perturbations that it
is equal to that of light's in vacuum, and this constraint
basically specifies the way that the scalar potential and the
non-minimal couplings of the theory can be chosen. The whole
theoretical framework, which belongs to the larger class of
Horndeski theories, cannot yield viable results, due to the fact
that the primordial gravitational wave speed is not equal to that
of light's. Thus we study this theory by also imposing the
constraint of having gravity wave speed equal to the light speed.
We directly examine the inflationary phenomenology of our
theoretical framework and by assuming the slow-roll conditions, we
derived the equations of motion in such a way so that analytical
results may be extracted. By using several well motivated models
we demonstrate the framework leads to a viable phenomenology.
\end{abstract}


\maketitle

\section{Introduction}

The striking GW170817 event \cite{TheLIGOScientific:2017qsa} of
the merging of two neutron stars has affected significantly the
perspective of theoretical cosmologists for modified gravity
theories. This is due to the fact that the GW170817 event was
accompanied by a kilonova event the GRB170817A \cite{GBM:2017lvd},
with the electromagnetic radiation arriving almost simultaneously
with the gravitational wave. This event immediately erased from
the stage of viable cosmological and astrophysical theories, all
the theories which predict a gravitational wave speed different
from that of light's, see Refs.
\cite{Ezquiaga:2017ekz,Baker:2017hug,Creminelli:2017sry,Sakstein:2017xjx}
for an account on this topic. In cosmology, this affected quite a
number of well-motivated inflationary theories which belong to a
large category of Horndeski theories
\cite{horndeskioriginal,Kobayashi:2019hrl,Kobayashi:2016xpl,Crisostomi:2016tcp,Bellini:2015xja,Gleyzes:2014qga,Lin:2014jga,Deffayet:2013lga,Bettoni:2013diz,Koyama:2013paa,Starobinsky:2016kua,Capozziello:2018gms,BenAchour:2016fzp,Starobinsky:2019xdp},
and specifically the non-minimal kinetic coupling theories
\cite{Sushkov:2009hk,Minamitsuji:2013ura,Saridakis:2010mf,Barreira:2013jma,Sushkov:2012za,Barreira:2012kk,Skugoreva:2013ooa,Gubitosi:2011sg,Matsumoto:2015hua,Deffayet:2010qz,Granda:2010hb,Matsumoto:2017gnx,Gao:2010vr,Granda:2009fh,Germani:2010gm,Fu:2019ttf}.
The effective inflationary Lagrangian is a puzzle for modern
theoretical physicists and cosmologists, and one well-motivated
candidate for the effective inflationary Lagrangian is the string
motivated Einstein-Gauss-Bonnet theory
\cite{Hwang:2005hb,Nojiri:2006je,Cognola:2006sp,Nojiri:2005vv,Nojiri:2005jg,Satoh:2007gn,Bamba:2014zoa,Yi:2018gse,Guo:2009uk,Guo:2010jr,Jiang:2013gza,Kanti:2015pda,vandeBruck:2017voa,Kanti:1998jd,Pozdeeva:2020apf,Fomin:2020hfh,DeLaurentis:2015fea,Chervon:2019sey,Nozari:2017rta,Odintsov:2018zhw,Kawai:1998ab,Yi:2018dhl,vandeBruck:2016xvt,Kleihaus:2019rbg,Bakopoulos:2019tvc,Maeda:2011zn,Bakopoulos:2020dfg,Ai:2020peo,Odintsov:2019clh,Oikonomou:2020oil,Odintsov:2020xji,Oikonomou:2020sij,Odintsov:2020zkl,Odintsov:2020sqy,Odintsov:2020mkz,Easther:1996yd,Antoniadis:1993jc,Antoniadis:1990uu,Kanti:1995vq,Kanti:1997br}.
The Einstein-Gauss-Bonnet theories are appealing because the
Lagrangian consists of a canonical scalar field part and of one
non-minimal coupling of the scalar field to the Gauss-Bonnet
invariant, and these theories lead to second order field
equations. However, Einstein-Gauss-Bonnet theories are affected by
the GW170817 event, since they predict a primordial tensor
perturbation propagation speed different from that of light's. In
some previous works
\cite{Odintsov:2019clh,Oikonomou:2020oil,Odintsov:2020xji,Oikonomou:2020sij,Odintsov:2020zkl,Odintsov:2020sqy,Odintsov:2020mkz},
we developed a theoretical framework which effectively solved the
problem of the primordial gravitational wave speed for
Einstein-Gauss-Bonnet theories. The framework was based on making
the primordial gravitational wave speed equal to that of light's
explicitly, and this constraint resulted on a constraint
differential equation that the non-minimal coupling of the scalar
field to the Gauss-Bonnet invariant must satisfy. In conjunction
with the slow-roll conditions, we found a particularly simple set
of field equations, with the striking new feature being that the
non-minimal coupling of the scalar field with the Gauss-Bonnet
invariant, and the scalar potential, must satisfy a differential
equation of a particular form. Therefore, these two functions must
not be arbitrarily chosen, but obey a specific differential
equation. This feature is entirely new, since in the existing
literature these two functions are arbitrarily chosen.

In this work, we shall extend the theoretical framework of Ref.
\cite{Oikonomou:2020sij}, including a non-minimal coupling of the
scalar field on the Ricci scalar. As we show, this new feature
enables one to have more freedom on producing a viable
inflationary phenomenology. The purpose of this work is to
demonstrate that the kinetic coupling corrected non-minimally
coupled Einstein-Gauss-Bonnet theory can produce a viable
inflationary phenomenology compatible with the latest Planck
constraints on inflation \cite{Akrami:2018odb}. By exploiting
several appropriately chosen models, and by using several well
motivated slow-roll compatible approximations, we demonstrate that
the theoretical framework of non-minimally coupled
Einstein-Gauss-Bonnet theory can produce a viable inflationary
phenomenology and at the same time it can also be compatible with
the GW170817 event, since the propagation speed of the tensor
perturbations, the primordial gravitations wave speed, is equal to
that of light's.

Before getting to  the core of this study, let us discuss an
important issue having to do with the motivation to study such
extended forms of modified gravities. We live in the era of
precision cosmology, so in principle many models seem to be viable
and compatible with the Planck 2018 data, like for example the
Starobinsky model and so on. Then why not sticking with the
Starobinsky description, and instead studying more difficult
models. The reason is simple. In fifteen years from now the LISA
collaboration will start to deliver the first data on primordial
gravitational waves. Thus LISA will definitely answer if inflation
took place or not. This is a crucial point, scalar field theories,
like the Starobinsky model in the Einstein frame or its Jordan
frame version the $R^2$ model, predict quite low power spectrum of
primordial gravitational waves, lower compared to the
sensitivities of LISA. Thus if LISA actually verifies a signal of
primordial gravitational waves, one should be sure where it comes
from. Is this due to a low reheating temperature and how low. Both
scalar field theories and $f(R)$ gravity may have an enhanced
signal of primordial gravitational waves, for a sufficiently low
reheating temperature. However, they also need a blue tilted
tensor spectral index, and this is impossible for these theories.
On the other hand Einstein-Gauss-Bonnet theories can yield easily
a blue tilted tensor spectral index. Thus theorists in the next
decade must be highly prepared for a plethora of possibilities
revealed by the LISA collaboration or other similar experiments,
like the Square Kilometer Array. This is the main motivation, to
know in the best way what phenomenological implications have the
available theoretical models existing. This is the main engine
that powers science. The experiments will verify if a theoretical
framework is consistent or not, like the LHC experiment actually
did with many theoretical approaches.

Another motivation for studying extended Einstein-Gauss-Bonnet
theories, is that these theories are basically string corrections,
and since inflation occurs chronologically quite close to the
Planck era, it might be possible that such string corrections
actually appear in the effectively inflationary Lagrangian.

\section{Aspects of Non minimally coupled/Kinetic Coupled Einstein-Gauss-Bonnet Gravity}

In the present paper, we shall study the dynamics of
Einstein-Gauss-Bonnet gravity in the presence of a non-trivial
kinetic coupling, and with a non-minimally coupling of the scalar
field to the Ricci scalar. This combination of string corrections
is one of the cases introduced in \cite{Hwang:2005hb}, and one of
the cases that can yield a massless primordial graviton and at the
same time can be studied in an analytic way under the slow-roll
assumption. The gravitational action of the non-minimal
Einstein-Gauss-Bonnet kinetic coupling corrected gravity is,
\begin{equation}
\centering
\label{action}
S=\int{d^4x\sqrt{-g}\left(\frac{h(\phi)R}{2\kappa^2}-\frac{\omega}{2}g^{\mu\nu}\partial_\mu\phi\partial_\nu\phi-V(\phi)-\xi(\phi)\left(\mathcal{G}+cG^{\mu\nu}\partial_\mu\phi\partial_\nu\phi\right)\right)}\, ,
\end{equation}
where $g$ is the determinant of the metric tensor $g^{\mu\nu}$,
$R$ denotes the Ricci scalar, $ h(\phi)$ is the dimensionless
scalar non-minimal coupling, and $V(\phi)$ is the scalar
potential. Also, $c$ is the dimensionful coupling of the kinetic
coupling term $\xi (\phi)
G^{\mu\nu}\partial_\mu\phi\partial_\nu\phi$, with mass dimensions
$[m]^{-2}$. Moreover $\omega$ will be assumed to be equal to unity
at the end, but we keep it as it is the following, in order to
keep the most general forms of the equations and the resulting
inflationary phenomenology, in case the reader is interested in
analyzing the inflationary phenomenology in the phantom scalar
case. Furthermore, $\xi(\phi)$ in the Gauss-Bonnet scalar function
which in this case is coupled to both the Gauss-Bonnet topological
invariant $\mathcal{G}$ and to Einstein's tensor
$G^{\mu\nu}=R^{\mu\nu}-\frac{1}{2}g^{\mu\nu}R$, via the kinetic
coupling term $\sim G^{\mu\nu}\partial_\mu\phi\partial_\nu\phi$,
where
$\mathcal{G}=R_{\mu\nu\sigma\rho}R^{\mu\nu\sigma\rho}-4R_{\mu\nu}R^{\mu\nu}+R^2$,
with $R_{\mu\nu\sigma\rho}$ and $R_{\mu\nu}$ being the Riemann
curvature tensor and the Ricci tensor respectively. This
particular will prove to be of paramount important in subsequent
calculations. Concerning the cosmological background, throughout
this paper we shall assume that it is described by a flat
Friedman-Robertson-Walker metric with the line element being,
\begin{equation}
\centering
\label{metric}
ds^2=-dt^2+a^2(t)\delta_{ij}dx^idx^j\, ,
\end{equation}
where $a(t)$ is the scale factor as usual. For this background,
the Ricci scalar and the Gauss-Bonnet topological invariant in
terms of Hubble's parameter are $H=\frac{\dot a}{a}$ as
$R=6(2H^2+\dot H)$ and $\mathcal{G}=24H^2(H^2+\dot H)$
respectively, where the ``dot'' as usual implies differentiation
with respect to cosmic time $t$. As a final step, we shall assume
that the scalar field is homogeneous, a valid assumption inspired
from the line element, which facilitates our study as well since
now the kinetic term of the scalar field takes the form
$-\frac{\omega}{2}\dot\phi^2$. Before we proceed with the
equations of motion and the overall phenomenology, we shall
consider the propagation velocity of the primordial gravitational
waves.

For the gravitational action (\ref{action}), due to the two string
corrections being present in the gravitational action, the
primordial gravitational waves propagate through spacetime with a
velocity which does not necessarily coincide with that of light's.
In fact, the general expression for the propagation velocity of
the tensor perturbations is \cite{Hwang:2005hb},
\begin{equation}
\centering
\label{cT}
c_T^2=1-\frac{Q_f}{2Q_t}\, ,
\end{equation}
where the auxiliary functions $Q_f$ and $Q_t$ are given by the
expressions $Q_f=16(\ddot\xi-H\dot\xi)+4c\xi\dot\phi^2$ and
$Q_t=\frac{h(\phi)}{\kappa^2}-8\dot\xi H+c\xi\dot\phi^2$.
Obviously, for all the cases that $c_T\neq 1$ in natural units,
the primordial gravitational waves, which are basically the tensor
perturbations, propagate in a different way in comparison to
electromagnetic waves, and the GW170817 merging event, which
involved a kilonova, the gravity waves came almost simultaneously
with the electromagnetic waves. Therefore, the propagation
velocities of gravitational and electromagnetic waves are quite
close in magnitude. As we explained in the introduction, there is
no reason that primordial gravitational waves should have a
propagation speed different from that of light's. In this line of
research, the compatibility with the GW170817 event can be
restored easily by equating the numerator of the second term with
zero, that is $Q_f=0$. Thus, the realization that the graviton is
massless forces the Gauss-Bonnet function $\xi(\phi)$ to satisfy
the differential equation $4(\ddot\xi-H\dot\xi)+c\xi\dot\phi^2=0$.
This was also the case with the minimally coupled gravity in Ref.
\cite{Oikonomou:2020sij}, since the only change is the different
form of $Q_t$ which does not affect the imposed condition
$c_T^2=1$. Thus, by rewriting the aforementioned relation is terms
of the scalar field and assuming that the slow-roll conditions for
the scalar field are valid, meaning that $\ddot\phi\ll H\dot\phi$,
then the following relation for the time derivative of the scalar
field is obtained,
\begin{equation}
\centering
\label{dotphi}
\dot\phi=\frac{4H\xi'}{4\xi''+c\xi}\, ,
\end{equation}
where ``prime'' denotes differentiation with respect to the scalar
field for simplicity. According to the gravitational action
(\ref{action}), the equations of motion can be extracted by simply
implementing the variation principle with respect to the scalar
field and the metric. Due to extra string contributions,
materialized by the kinetic coupling term, and of course due to
the extra non-minimal coupling, it is expected that their
respective form is quite lengthy. The equations of motion read,
\begin{equation}
\centering
\label{motion1}
\frac{3hH^2}{\kappa^2}=\frac{1}{2}\omega\dot\phi^2+V-\frac{3H\dot h}{\kappa^2}+24\dot\xi H^3-9c\xi H^2\dot\phi^2\, ,
\end{equation}
\begin{equation}
\centering
\label{motion2}
-\frac{2h\dot H}{\kappa^2}=\omega\dot\phi^2+\frac{h''\dot\phi^2+h'\ddot\phi-H\dot h}{\kappa^2}-8H^2(\ddot\xi-H\dot\xi)-16\dot\xi H\dot H+c\dot\phi\left(2\xi(\dot H-3H^2)\dot\phi+4H\xi\ddot\phi+2H\dot\xi\dot\phi\right)\, ,
\end{equation}
\begin{equation}
\centering \label{motion3}
V'+\omega(\ddot\phi+3H\dot\phi)-\frac{Rh'}{2\kappa^2}+\xi'\mathcal{G}-3c\left(H^2(\dot\xi\dot
H+2\xi\ddot\phi)+2H(2\dot H+3H^2)\xi\dot\phi\right)=0\, .
\end{equation}
As it can easily be inferred, the equations of motion are quite
perplexed and an analytical solution cannot be extracted easily,
unless the slow-roll conditions are invoked, in the same way as in
the minimally coupled scalar theory of inflation. One
theoretically optimal approach would be to solve the differential
equations (\ref{motion1})-(\ref{motion3}) numerically. This task
however is formidable for three main reasons: Firstly there is no
specific motivation to choose in a specific way the initial
conditions for the scalar field and for the Hubble rate. One for
example may choose at the first horizon crossing the initial
conditions to be $\phi(t_i)\sim M_p$ and $H(t_i)$ to be a de
Sitter or a quasi-de Sitter vacuum, but this is arbitrary. One may
assign some other initial value to the Hubble rate which may still
lead to an inflationary evolution for example of the form
$H(t)\sim 1/t$, but this again would be a different numerical
solution. Secondly and more importantly, even if we were able to
choose one of the many different initial conditions, each
corresponding to different numerical solutions, one cannot be sure
whether to apply the initial conditions at the end of the
inflationary era and solve backwards the differential equations.
For example if one is sure about the reheating era, then we can
fix the initial conditions at the beginning of reheating, which
occurs at the end of inflation. But the reheating era is
speculative, as is inflation itself. At late-time one is sure
about what is going on, the matter era is followed by a dark
energy era, so the initial conditions can be physically motivated.
Inflation is different. Thirdly, even if we found a way to
determine the initial conditions, the results that would be
obtained by solving numerically Eqs.
(\ref{motion1})-(\ref{motion3}) would correspond to a
non-slow-roll solution. There is no practical way to quantify the
slow-roll condition. The same problems described above occur for
all inflationary theories, especially the ones involving a scalar
field. The initial condition problem overwhelms inflation, this is
why we reside to the more practical slow-roll assumption and the
corresponding semi-analytic analysis. Even in the simple scalar
field theory, the numerical solution obtained by solving the
non-slow-roll equations of motion, correspond to an entirely
different solution compared to the slow-roll solution. This is why
in simple scalar field theory, attractors are investigated if they
exist, always under the slow-roll assumption. This problem would
be difficult to tackle for the case at hand, but still it is an
interesting future perspective.

We shall make the exact same approximations as in the minimally
coupled scalar field theory case, the slow-roll approximations,
$\dot H\ll H^2$, $\frac{1}{2}\omega\dot\phi^2\ll V$ and
$\ddot\phi\ll H\dot\phi$, which are reasonable assumptions since
Eq. (\ref{dotphi}) was essentially produced by implementing the
third relation. Depending on the choice of the Gauss-Bonnet
function $\xi(\phi)$ and the Ricci scalar coupling functions as
well, there exist various approximations that can be used in order
to simplify the analytical study of the inflationary
phenomenology. We review these in the last section of the article
and by choosing one of these, the equation of motion are
simplified as follows,
\begin{equation}
\centering
\label{motion4}
\frac{3hH^2}{\kappa^2}=V-\frac{12H^2 h'\xi'}{\kappa^2(4\xi''+c\xi)}\, ,
\end{equation}
\begin{equation}
\centering
\label{motion5}
-\frac{2h\dot H}{\kappa^2}=H^2\frac{4\xi'}{4\xi''+c\xi}\left(\left(\frac{h''}{\kappa^2}+\omega\right)\frac{4\xi'}{4\xi''+c\xi}-\frac{h'}{\kappa^2}\right)\, ,
\end{equation}
\begin{equation}
\centering \label{motion6} V'+3
H^2\left(\frac{4\omega\xi'}{4\xi''+c\xi}-2\frac{h'}{\kappa^2}\right)=0\,
.
\end{equation}
Comparing the two forms of the equations of motion respectively,
it is clear that these reasonable assumptions facilitate our study
greatly. Also, as was the case with the minimally coupled scalar
field case, Eq. (\ref{motion5}) specifies the first slow-roll
index as we shall demonstrate in subsequent calculations, thus
simplifying this particular index, since every quantity can be
written as a function of the scalar field, is a priority.

In order to quantify the inflationary phenomenology study, we
shall calculate the slow-roll indices, and due to the string
corrections, we have six slow-roll indices as shown below,
\begin{align}
\centering
\label{slowroll}
\epsilon_1&=-\frac{\dot H}{H^2}&\epsilon_2&=\frac{\ddot\phi}{H\dot\phi}&\epsilon_3&=\frac{\dot h}{2Hh}&\epsilon_4&=\frac{\dot E}{2HE}&\epsilon_5&=\frac{\dot F+Q_a}{2HQ_t}&\epsilon_6&=\frac{\dot Q_t}{2HQ_t}\, ,
\end{align}
where $F=\frac{h}{\kappa^2}$, $Q_a=-8\dot\xi
H^2+4c\xi\dot\phi^2H$,
$E=\frac{F}{\dot\phi^2}\left(\omega\dot\phi^2+\frac{3(\dot
F+Q_a)^2}{2Q_t}+Q_c\right)$ and $Q_c=-6c\xi\dot\phi^2 H^2$. Owing
to the fact that compatibility with the GW170817 event specifies
the form of $\dot\phi$, the slow-roll indices in this case along
with the auxiliary parameters used, take the following forms,
\begin{equation}
\centering
\label{index1}
\epsilon_1=\frac{2\xi'}{4\xi''+c\xi}\left(\left(\frac{h''}{h}+\frac{\kappa^2\omega}{h}\right)\frac{4\xi'}{4\xi''+c\xi}-\frac{h'}{h}\right)\, ,
\end{equation}
\begin{equation}
\centering
\label{index2}
\epsilon_2=\frac{4\xi''}{4\xi''+c\xi}-\epsilon_1-\frac{4\xi'(4\xi'''+c\xi')}{(4\xi''+c\xi)^2}\, ,
\end{equation}
\begin{equation}
\centering
\label{index3}
\epsilon_3=\frac{2\xi'}{4\xi''+c\xi}\frac{h'}{h}\, ,
\end{equation}
\begin{equation}
\centering
\label{index4}
\epsilon_4=\frac{2\xi'}{4\xi''+c\xi}\frac{E'}{E}\, ,
\end{equation}
\begin{equation}
\centering
\label{index5}
\epsilon_5=\frac{1}{Q_t}\left(\epsilon_3 F+\frac{Q_a}{2}\right)\, ,
\end{equation}
\begin{equation}
\centering
\label{index6}
\epsilon_6=\frac{2\xi'}{4\xi''+c\xi}\frac{Q_t'}{Q_t}\, ,
\end{equation}
and moreover, the auxiliary functions $Q_a$, $Q_c$, $Q_d$ and
$Q_e$ used above, these are,
\begin{equation}
\centering
Q_a=-\frac{32\xi'^2}{4\xi''+c\xi} H^3+4c\xi H^3\left(\frac{4\xi'}{4\xi''+c\xi}\right)^2\, ,
\end{equation}
\begin{equation}
\centering
\label{Qc}
Q_c=-6c\xi H^4\left(\frac{4\xi'}{4\xi''+c\xi}\right)^2\, ,
\end{equation}
\begin{equation}
\centering
\label{Qd}
Q_d=-4c\xi H^2\dot H\left(\frac{4\xi'}{4\xi''+c\xi}\right)^2\, ,
\end{equation}
\begin{equation}
\centering
\label{Qe}
Q_e=-32\frac{4\xi'^2}{4\xi''+c\xi}H\dot H+32c\frac{\xi'^2}{(4\xi''+c\xi)^2}H^3\left(\frac{4\xi'^2}{4\xi''+c\xi}+2\xi(\epsilon_2-1)\right)\, ,
\end{equation}
where $Q_d$ and $Q_e$ are introduced for later convenience. In
this case, it was deemed suitable to write indices $\epsilon_5$
and $\epsilon_6$ in this manner due to the fact that the
non-minimally coupling to the Ricci scalar, makes the indices
quite perplexed or lengthy, hence in order to avoid this we simply
express them in terms of the scalar field. The slow-roll indices
are also connected to the observational indices, namely the
spectral index of primordial scalar curvature perturbations $n_S$,
the tensor spectral index $n_T$, and the tensor-to-scalar ratio
$r$, in the following way,
\begin{align}
\centering
\label{observed}
n_S&=1-2\frac{2\epsilon_1+\epsilon_2-\epsilon_3+\epsilon_4}{1-\epsilon_1}&n_T&=-2\frac{\epsilon_1+\epsilon_6}{1-\epsilon_1}&r&=16\left|\left(\epsilon_1+\epsilon_3+\frac{\kappa^2}{4hH^2}(2Q_c+Q_d-HQ_e)\right)\frac{hc_A^3}{\kappa^2Q_t}\right|\, ,
\end{align}
where $c_A$ denotes the sound wave velocity specified as,
\begin{equation}
\centering
\label{cA}
c_A^2=1+\frac{2Q_tQ_d+(\dot F+Q_a)Q_e}{2\omega Q_t\dot\phi^2+3(\dot F+Q_a)^2+2Q_tQ_c}\, ,
\end{equation}
As a final step, we introduce the new form of the $e$-foldings
number, depending solely on the Gauss-Bonnet scalar coupling
function. Since $N=\int_{t_i}^{t_f}{Hdt}$ where $t_f-t_i$
signifies the duration of the inflationary era, and due to the
fact that $\frac{d}{dt}=\dot\phi\frac{d}{d\phi}$, we have,
\begin{equation}
\centering \label{efolds}
N=\int_{\phi_i}^{\phi_f}{\frac{4\xi''+c\xi}{4\xi'}d\phi}\, .
\end{equation}
Having the above relations at hand, we can proceed in the next
section by examining explicitly the inflationary phenomenology for
some models of interest.

\section{Inflationary Phenomenology of Specific Models}

In this section we shall demonstrate that the non-minimally
coupled Einstein-Gauss-Bonnet theory with non-minimal kinetic term
can produce a viable inflationary phenomenology. We shall use
several models of interest and we shall investigate under which
conditions these models can yield a viable inflationary era.

\subsection{Exponential Gauss-Bonnet And Linear Ricci Coupling}

We begin our study with one of the most convenient choices for the
coupling functions. Let,
\begin{equation}
\centering
\label{xiA}
\xi(\phi)=\lambda_1e^{\gamma_1\kappa\phi}\, ,
\end{equation}
\begin{equation}
\centering
\label{hA}
h(\phi)=\Lambda_1\kappa\phi\, ,
\end{equation}
where $\lambda_1$, $\gamma_1$ and $\Lambda_1$ are the free
parameters of the model. This model is capable of producing
results compatible with the observations assuming that the only
string corrections in action (\ref{action}) are
$\xi(\phi)\mathcal{G}$ thus it is interesting to examine the
possibility of viability in the presence of extra string
corrections. In addition, since the Ricci coupling is linear, it
turns out that $h''=0$ hence Eq. (\ref{motion5}) is simplified
without the need of any assumption. For the time being, we shall
make use of the following equations of motion,
\begin{equation}
\centering
\label{motion1A}
H^2=\frac{\kappa^2V}{3h}\, ,
\end{equation}
\begin{equation}
\centering
\label{motion2A}
\dot H=-\frac{2H^2\xi'}{4\xi''+c\xi}\left(\frac{\kappa^2\omega}{h}\frac{4\xi'}{4\xi''+c\xi}-\frac{h'}{h}\right)\, ,
\end{equation}
\begin{equation}
\centering
\label{motion3A}
V'+3H^2\left(\frac{4\omega\xi'}{4\xi''+c\xi}-2\frac{h'}{\kappa^2}\right)=0\, ,
\end{equation}
In the minimally coupled case, the choice of an exponential
Gauss-Bonnet coupling led to a constant $\dot\phi$, hence the
reason it is chosen to be as in Eq. (\ref{xiA}), is because Eq.
(\ref{motion2A}) and subsequently the slow-roll index $\epsilon_1$
are simplified. Due to this choice, the scalar potential reads,
\begin{equation}
\centering
\label{VA}
V(\phi)=V_1\left(\frac{c\phi}{\kappa}+4\gamma_1^2\kappa\phi\right)^{\frac{2c\Lambda_1+4\gamma_1\kappa^2(2\gamma_1\Lambda_1-\omega)}{\Lambda_1(c+(2\gamma_1\kappa)^2)}}\, ,
\end{equation}
where $V_1$ is the integration constant with mass dimensions
$[m]^4$. In this case as well, the scalar potential is a power-law
model with specific exponent, which is not necessarily an integer
as usual. In the following we shall showcase that viability can be
achieved with the exponent being quite close to 2. Let us now
proceed with the slow-roll indices. Due to the coupling functions
only, we have,
\begin{equation}
\centering
\label{index1A}
\epsilon_1=-2\gamma_1\kappa\frac{c\Lambda_1+4\gamma_1\kappa^2(\gamma_1\Lambda_1-\omega)}{\Lambda_1(c+(2\gamma_1\kappa)^2)\phi}\, ,
\end{equation}
\begin{equation}
\centering
\label{index2A}
\epsilon_2=-\epsilon_1\, ,
\end{equation}
\begin{equation}
\centering
\label{index3A}
\epsilon_3=\frac{2\gamma_1\kappa}{((2\gamma_1\kappa)^2+c)\phi}\, ,
\end{equation}
In this case, only the first three slow-roll indices have elegant
forms while the rest have too lengthy expressions, so we omitted
their final form. Due to the fact that the Gauss-Bonnet coupling
is exponential, indices $\epsilon_1$ and $\epsilon_2$ are
opposite. In particular, index $\epsilon_1$ depends on $\phi$ with
an inverse power-law dependence, thus only a single field solution
exists. In consequence, letting $\epsilon_1$ become of order
$\mathcal{O}(1)$ and using Eq. (\ref{efolds}), produces the
following expressions for the scalar field during the first
horizon crossing and the final stage of inflation,
\begin{equation}
\centering
\label{phiiA}
\phi_i=\phi_f-\frac{4N\gamma_1\kappa}{(2\gamma_1\kappa)^2+c}\, ,
\end{equation}
\begin{equation}
\centering \label{phifA}
\phi_f=-2\gamma_1\frac{c\kappa\Lambda_1+4\gamma_1\kappa^3(\gamma_1\Lambda_1-\omega)}{\Lambda_1((2\gamma_1\kappa)^2+c)^2}\,
.
\end{equation}
This in turn implies that the scalar field has a unique evolution
throughout the inflationary era. Concerning the observational
indices, the results are produced by designating the free
parameters of the model. Assigning the values ($\omega$,
$\lambda_1$, $\Lambda_1$, $V_1$, $N$, $c$, $\gamma_1$)=(1, 1, 100,
1, 60, 0.002, -10) in reduced Planck units where $\kappa=1$, then
the scalar spectral index of primordial curvature perturbations
becomes $n_S=0.967296$, the tensor spectral index obtains the
value $n_T=-0.0000167$ and finally the tensor-to-scalar ratio is
equal to $r=0.00013223$, which are obviously compatible values
with the latest Planck 2018 data. Furthermore, the scalar field
seems to decrease with time as $\phi_i=6.05002$ and
$\phi_f=0.0500497$, the sound wave velocity $c_A$ is equal to
unity, as expected hence no ghost instabilities are present.
Furthermore, the numerical values of the slow-roll indices are
$\epsilon_1=0.00827266$, $\epsilon_2=-\epsilon_1$,
$\epsilon_3=-0.00826439$, $\epsilon_4=-0.00030204$ and finally,
indices $\epsilon_5$ and $\epsilon_6$ are both equal to
$\epsilon_6$. The small values of the slow-roll indices is
indicative of the validity of the slow-roll approximations imposed
previously. In Figs. 1 and 2 we present the dependence of the
observational indices on two of the free parameters of the model.
As it shown, the viability of the model is achieved for a wide
range of the free parameters.
\begin{figure}[h!]
\centering
\label{plot1}
\includegraphics[width=17pc]{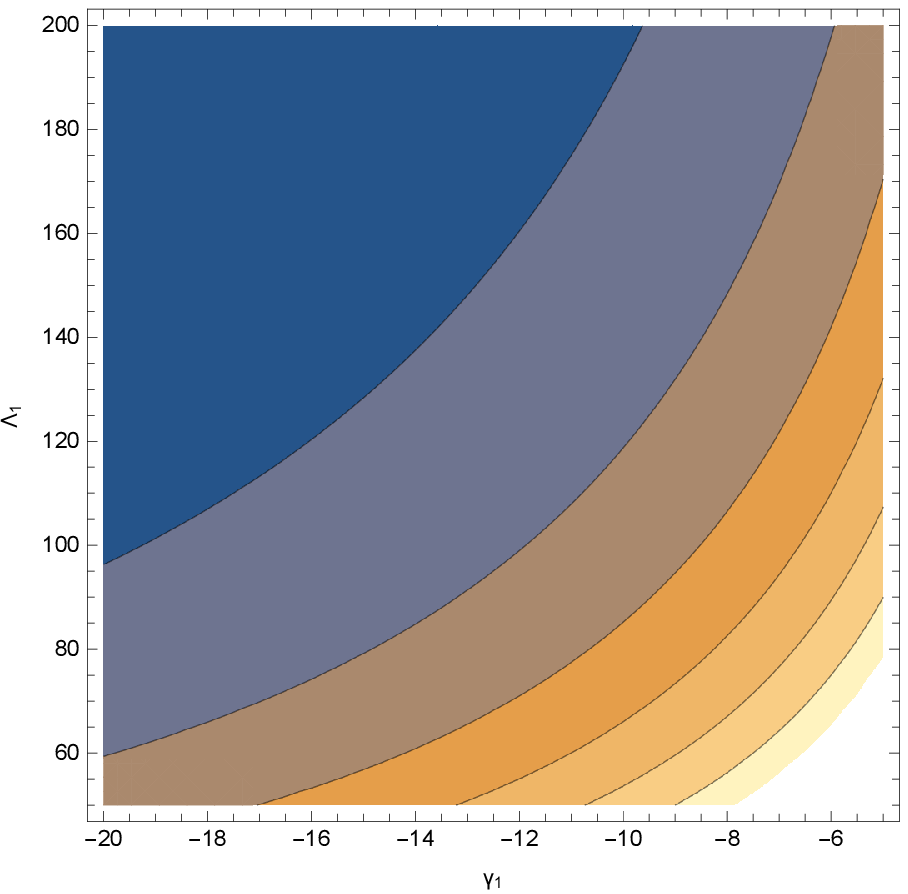}
\includegraphics[width=3pc]{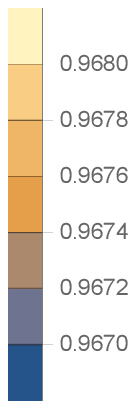}
\includegraphics[width=17pc]{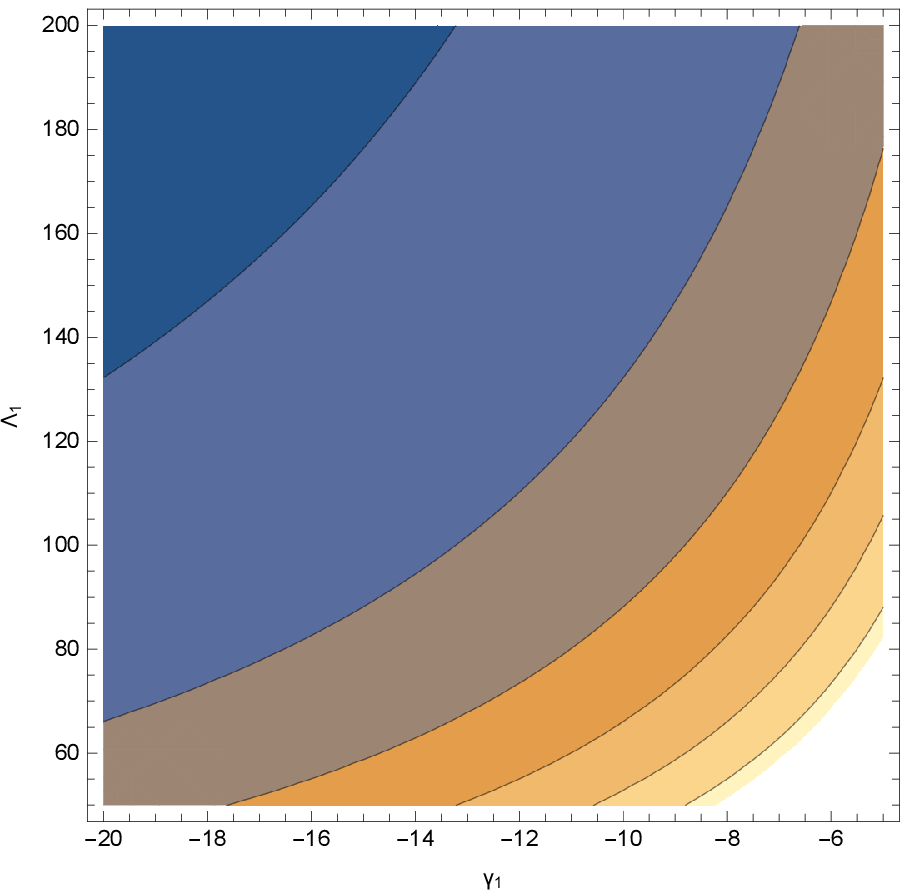}
\includegraphics[width=3pc]{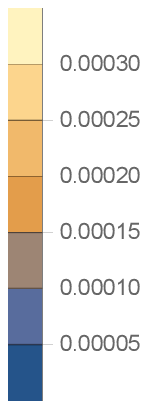}
\caption{The scalar spectral index of primordial curvature
perturbations $n_S$ (left) and tensor-to-scalar ratio $r$ (right)
as functions of $\gamma_1$ and $\Lambda_1$. Their values are
chosen in the ranges [-20,-5] and [50,200] respectively. The plots
are derived from the initial assumption where equations
(\ref{motion1A}) through (\ref{motion3A}) are valid.}
\end{figure}
There are many differences between the minimally and non-minimally
coupled case which are not attributed to this specific choice of
coupling functions. Firstly, the string corrections are now indeed
inferior compared to other terms and one does not need to decrease
$\lambda_1$ in order to make this assumption valid. In this case,
trivial values seem to yield a viable phenomenology. Moreover, the
non-minimally coupled term in stronger than string corrections, as
it can easily be inferred from the values of slow-roll indices
$\epsilon_3$, $\epsilon_5$ and $\epsilon_6$. As a result, indices
$\epsilon_5$ and $\epsilon_6$ are once again equal, as was the
case with the minimally and non-minimally coupled string
corrections of the form $\xi(\phi)\mathcal{G}$ only. As mentioned
before, this specific choice of values for the free parameters
leads to the exponent of the scalar potential being equal to
2.001, which is close to 2 as stated before. Hence, the overall
model is quite simple as it is comprised of two trivial power law
scalar functions and one exponential.

Finally, we mention that each and every approximation made
throughout this procedure is indeed valid. Beginning from the
slow-roll conditions, which here hinted to be indeed valid from
the numerical values of the slow-roll indices during the first
horizon crossing, we note that $\dot H\sim\mathcal{O}(10)$ whereas
$H^2\sim\mathcal{O}(10^3)$,
$\frac{1}{2}\omega\dot\phi^2\sim\mathcal{O}(10)$ while
$V\sim\mathcal{O}(10^6)$ and finally $\ddot\phi\sim\mathcal{O}(1)$
compared to $H\dot\phi\sim\mathcal{O}(100)$. For the string
corrections, we mention that $24\dot\xi
H^3\sim\mathcal{O}(10^{-18})$ and
$9c\xi\dot\phi^2H^2\sim\mathcal{O}(10^{-23})$ where
$V\sim\mathcal{O}(10^6)$, $16\dot\xi H\dot
H\sim\mathcal{O}(10^{-21})$,
$8H^2(\ddot\xi-H\dot\xi)\sim\mathcal{O}(10^{-21})$ and
$c\dot\phi\left(2\xi\dot\phi(\dot
H-3H^2)+4H\xi\ddot\phi+2H\dot\xi\dot\phi\right)\sim\mathcal{O}(10^{-24})$
compared to $\dot\phi^2\sim\mathcal{O}(10)$ and $H\dot
h\sim\mathcal{O}(10^4)$ and finally,
$\xi'\mathcal{G}\sim\mathcal{O}(10^{-17})$,
$3c\left(H^2(\dot\xi\dot H+2\xi\ddot\phi)+2H(2\dot
H+3H^2)\xi\dot\phi\right)\sim\mathcal{O}(10^{-22})$ in contrast to
$V'\sim\mathcal{O}(10^6)$, $6H^2h'\sim\mathcal{O}(10^6)$ and
$3\omega H\dot\phi\sim\mathcal{O}(10^3)$. Last but not least, for
Eq. (\ref{motion4}), we mention that neglecting $3H\dot h$ in
order to produce Eq. (\ref{motion1A}) is valid since
$V\sim\mathcal{O}(10^6)$ while $3H\dot h\sim\mathcal{O}(10^4)$.
Hence, all the approximations made are indeed valid.
\begin{figure}[h!]
\centering
\label{plot2}
\includegraphics[width=17pc]{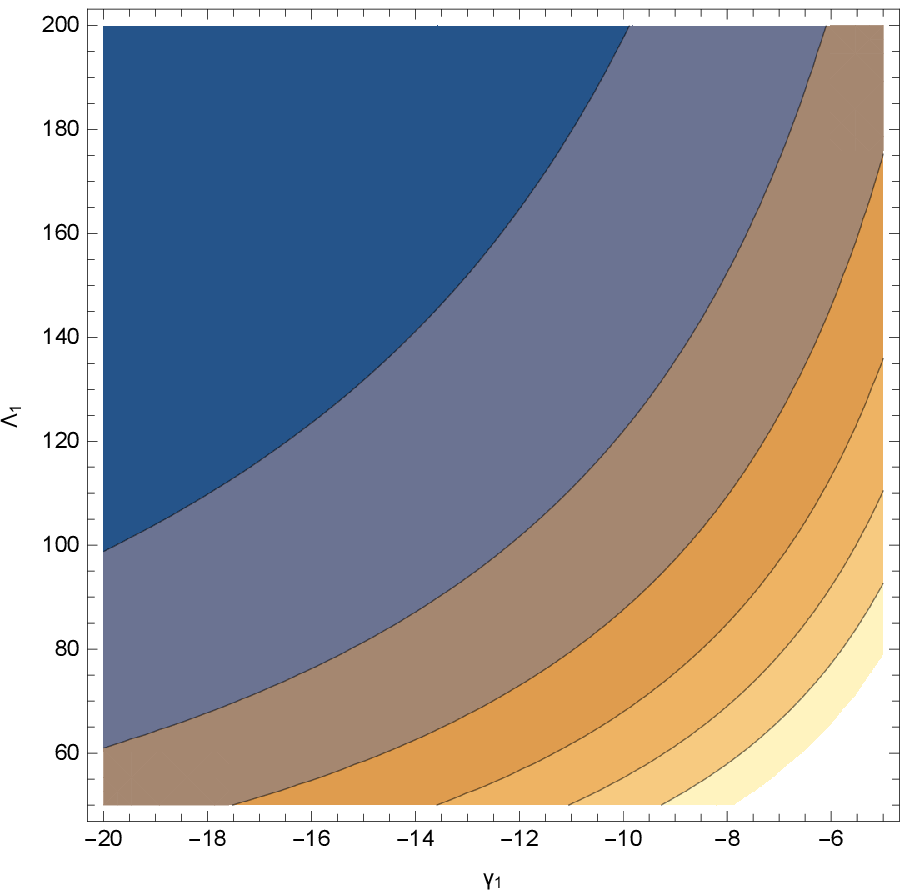}
\includegraphics[width=3pc]{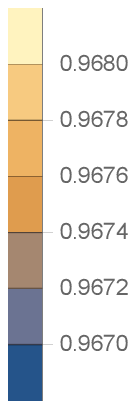}
\caption{Scalar spectral index of primordial curvature
perturbations $n_S$ depending on parameters $\gamma_1$ and
$\Lambda_1$ in the value ranges [-20,-5] and [50,200]
respectively. In the case of the only change being Eq.
(\ref{motion2A2}), the scalar spectral index remains the same.
Since the tensor-to-scalar ratio is effectively zero, no
corresponding plot is showcased.}
\end{figure}
As a last comment, it is worth mentioning that different equations
of motion are capable of producing similar results. As it was
hinted previously, since $3H\dot h$ is smaller than $V$, then
using Eq. (\ref{motion4}) as Hubble's form results in the exactly
same observational indices, without altering even a single one,
thus it is reasonable to discard such term. Moreover, $3\omega
H\dot\phi$ is orders of magnitude smaller than $V'$ which implies
that such term can be omitted. As a matter of fact, doing so leads
to an exponent for the scalar potential which is exactly 2. This
is the same regardless of keeping $3H\dot h$ or not in Eq.
(\ref{motion4}). The most striking result happens when
$\omega\dot\phi^2$ is neglected from Eq. (\ref{motion2}). Assuming
that the only change in the equations of motion is,
\begin{equation}
\centering
\label{motion2A2}
\dot H=\frac{2H^2\xi'}{4\xi''+c\xi}\frac{h'}{h}\, ,
\end{equation}
then for the same values for the free parameters, the scalar
spectral index experiences a mild change as now $n_S=0.967313$ but
on the other hand the tensor spectral index becomes equal to zero,
along with the tensor-to-scalar ratio which numerically speaking
is $r=3\cdot10^{-26}$ but essentially is zero. This approach
implies that no B (curl) modes are present, which is a striking
result compared to the one obtained previously, assuming that the
only change is in Hubble rate derivative. The same result is
acquired irrespective of $3H\dot h$ being in Eq. (\ref{motion4})
or $3H\dot\phi$ discarded from Eq. (\ref{motion3}). This could be
attributed to the choice of the free parameters but many other
pairs which were used do not seem to increase the tensor-to-scalar
ratio from the value $r=10^{-18}$ without breaking the value of
the scalar spectral index.


\subsection{Exponential Gauss-Bonnet And Power-Law Coupling}

Let us now study a similar model but with a different approach.
Instead of using a linear Ricci coupling, we shall assume a
general power-law model in order to obtain a nonzero $h''$
derivative. Suppose that,
\begin{equation}
\centering
\label{xiB}
\xi(\phi)=\lambda_2 e^{\gamma_2\kappa\phi}\, ,
\end{equation}
\begin{equation}
\centering
\label{hB}
h(\phi)=\Lambda_2(\kappa\phi)^m\, ,
\end{equation}
where once again $\lambda_2$, $\gamma_2$ and $\Lambda_2$ are the
free parameters of the model, but the subscript is changed in
order to differentiate from the previous model. Since this choice
was made in order to have $h''\neq0$, we shall take advantage of
such feature and see whether a viable phenomenology can be
produced. In the following, we shall assume that the proper
equations of motion are,
\begin{equation}
\centering
\label{motion1B}
H^2=\frac{\kappa^2V}{3h}\, ,
\end{equation}
\begin{equation}
\centering
\label{motion2B}
\dot H=-\frac{2H^2\xi'}{4\xi''+c\xi}\left(\frac{h''}{h}\frac{4\xi'}{4\xi''+c\xi}-\frac{h'}{h}\right)\, ,
\end{equation}
\begin{equation}
\centering
\label{motion3B}
V'+3H^2\left(\frac{4\omega\xi'}{4\xi''+c\xi}-2\frac{h'}{\kappa^2}\right)=0\, ,
\end{equation}
where we made certain assumptions and we kept the leading order
terms. In the end, we shall make comparisons between the different
assumptions which can be made, but for now we shall continue with
these. Due to this choice, the scalar potential reads,
\begin{equation}
\centering
\label{VB}
V(\phi)=V_2(\kappa\phi)^{2m}e^{\frac{4\gamma_1\kappa^2\omega}{(m-1)\Lambda_1((2\gamma_1\kappa)^2+c)}(\kappa\phi)^{1-m}}\, ,
\end{equation}
where now the integration constant is denoted as $V_2$. Here, the
scalar potential is a combination of a power-law and an
exponential function. In particular, the exponential has the form
of $\delta (\kappa\phi)^{1-m}$ with $\delta$ having a specific
form proportional to the rest free parameters. Hence, the
potential is a combination of the coupling functions. However, in
the case of $3\omega H\dot\phi$ being inferior to $V'$ in Eq.
(\ref{motion3B}), the exponential part disappears and the
potential is once again a power-law form. The same result is
derived in the case we studied in the previous subsection, with a
linear Ricci coupling, but now replacing $m=1$ is forbidden. In
addition, the slow-roll indices, or at least the first three, are,
\begin{equation}
\centering
\label{index1B}
\epsilon_1= -2m\gamma_2\kappa\frac{((2\gamma_2\kappa)^2+c)\phi-4(m-1)\gamma_2\kappa}{((2\gamma_2\kappa)^2+c)^2\phi^2}\, ,
\end{equation}
\begin{equation}
\centering
\label{index2B}
\epsilon_2=-\epsilon_1\, ,
\end{equation}
\begin{equation}
\centering
\label{index3B}
\epsilon_3=\frac{2m\gamma_2\kappa}{((2\gamma_2\kappa)^2+c)\phi}\, ,
\end{equation}
Once again, the rest of the slow-roll indices have quite lengthy
final form, so we omitted them, however it is worth mentioning
that this was not the case with the $\xi(\phi)\mathcal{G}$ string
corrections for $\epsilon_5$ and $\epsilon_6$ at least. Similarly
to the previous model, the initial and final value of the scalar
field are given by the following expressions,
\begin{equation}
\centering
\label{phiiB}
\phi_i=\phi_f-\frac{4N\gamma_2\kappa}{(2\gamma_2\kappa)^2+c}\, ,
\end{equation}
\begin{equation}
\centering \label{phifB} \phi_f=\frac{\pm\sqrt{\left(-2 c\gamma_2
\kappa  m-8\gamma_2^3 \kappa ^3 m\right)^2-4 \left(-c^2-8
c\gamma_2^2 \kappa ^2-16 \gamma_2^4 \kappa ^4\right)
\left(8\gamma_2^2 \kappa ^2 m^2-8\gamma_2^2 \kappa ^2 m\right)}-2
c\gamma_2\kappa  m-8 \gamma_2^3 \kappa ^3 m}{2 \left(c^2+8 c
\gamma_2^2 \kappa ^2+16 \gamma_2^4 \kappa ^4\right)}\, .
\end{equation}
Since a different equation for Hubble's derivative is used, we now
have two possible forms for the final value of the scalar field,
however we shall use only the positive. Letting ($\omega$,
$\lambda_2$, $\Lambda_2$, $V_2$, $N$, $c$, $\gamma_2$, $m$)=(1, 1,
100, 1, 60, 0.002, 30, 2) yields viable results since
$n_S=0.965156$, $n_T=-0.00057706$, and $r=0.00453744$ which are
compatible with the data provided by the Planck 2018 collaboration
\cite{Akrami:2018odb}. Moreover, $\epsilon_1=0.0171237$,
$\epsilon_3=-0.0168401$ and indices $\epsilon_4$ through
$\epsilon_6$ are equal to $\epsilon_3$.
\begin{figure}[h!]
\centering
\label{plot3}
\includegraphics[width=17pc]{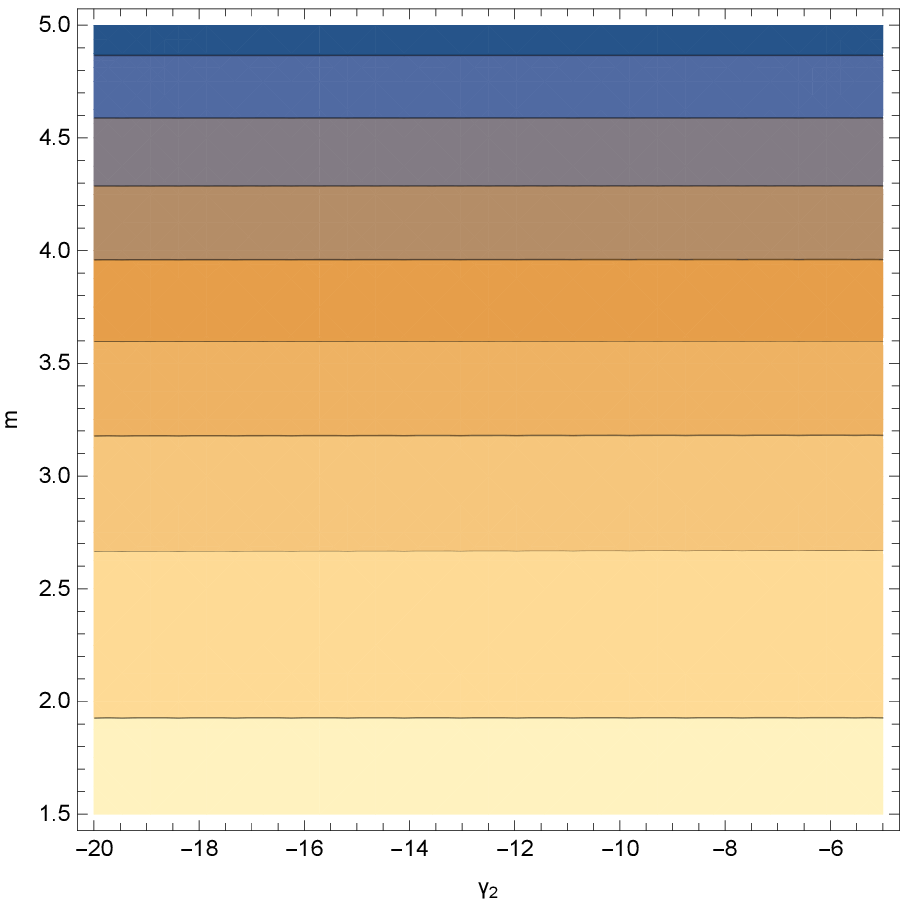}
\includegraphics[width=3pc]{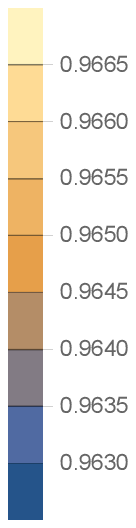}
\includegraphics[width=17pc]{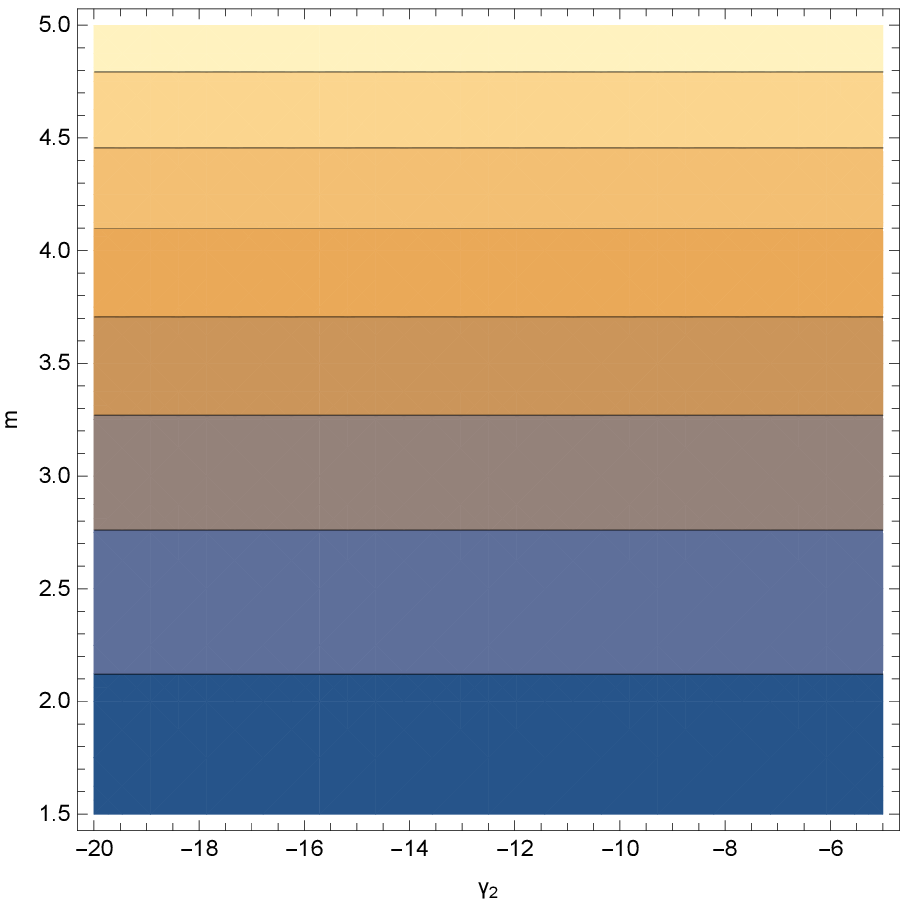}
\includegraphics[width=3pc]{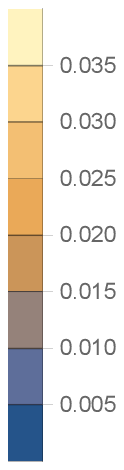}
\caption{Scalar spectral index of primordial curvature
perturbations $n_S$ (left) and tensor-to-scalar ratio $r$ (right)
depending on parameters $\gamma_2$ and $m$ ranging  from [-20,-5]
and [1.5,5] respectively. It becomes apparent that the dominant
contributor is exponent $m$ which seems to have different effects
on the observed indices. These plots correspond to the initial
equations of motion of this particular model.}
\end{figure}
In this case, the scalar potential has an appealing form. The
power-law part is $V\sim(\kappa\phi)^4$ whereas the exponential
part is written as $V\sim e^{\frac{1}{3.3\cdot10^{-4}\phi}}$,
implying that the exponential part is not so dominant compared to
the power-law. The behavior of the observational indices as
functions of several free parameters is depicted in Figs. 3 and 4.
As it can be seen, the phenomenological viability of the model is
guaranteed for a wide range of the values of the free parameters.

It is worth mentioning that this model is also valid, as all the
approximations made indeed apply. Concerning the slow-roll
indices, we have $\dot H\sim\mathcal{O}(10^{-4})$ whereas
$H^2\sim\mathcal{O}(10^{-2})$,
$\frac{1}{2}\omega\dot\phi^2\sim\mathcal{O}(10^{-6})$ in contrast
to $V\sim\mathcal{O}(10)$ and lastly,
$\ddot\phi\sim\mathcal{O}(10^{-6})$ and
$H\dot\sim\mathcal{O}(10^{-4})$ thus the slow-roll approximations
apply. For the string corrections, we have $24\dot\xi
H^3\sim\mathcal{O}(10^{-33}$ and
$9c\xi\dot\phi^2H^2\sim\mathcal{O}(10^{-39}$, $16\dot\xi H\dot
H\sim\mathcal{O}(10^{-35})$,
$8H^2(\ddot\xi-H\dot\xi)\sim\mathcal{O}(10^{-35})$ and
$c\dot\phi\left(2\xi\dot\phi(\dot
H-3H^2)+4H\xi\ddot\phi+2H\dot\xi\dot\phi\right)\sim\mathcal{O}(10^{-39})$,
$\xi'\mathcal{G}\sim\mathcal{O}(10^{-31})$ and
$3c\left(H^2(\dot\xi\dot H+2\xi\ddot\phi)+2H(2\dot
H+3H^2)\xi\dot\phi\right)\sim\mathcal{O}(10^{-37})$ thus string
corrections are negligible since in Planck units are effectively
zero. Finally, for equations (\ref{motion1B}) through
(\ref{motion3B}), we mention that $3H\dot
h\sim\mathcal{O}(10^{-1})$ while $V\sim\mathcal{O}(10)$,
$\omega\dot\phi^2\sim\mathcal{O}(10^{-5})$ while
$h''\dot\phi^2\sim\mathcal{O}(10^{-3})$ and $H\dot
h\sim\mathcal{O}(10^{-1})$ hence all the approximations are indeed
valid.
\begin{figure}[h!]
\centering
\label{plot4}
\includegraphics[width=17pc]{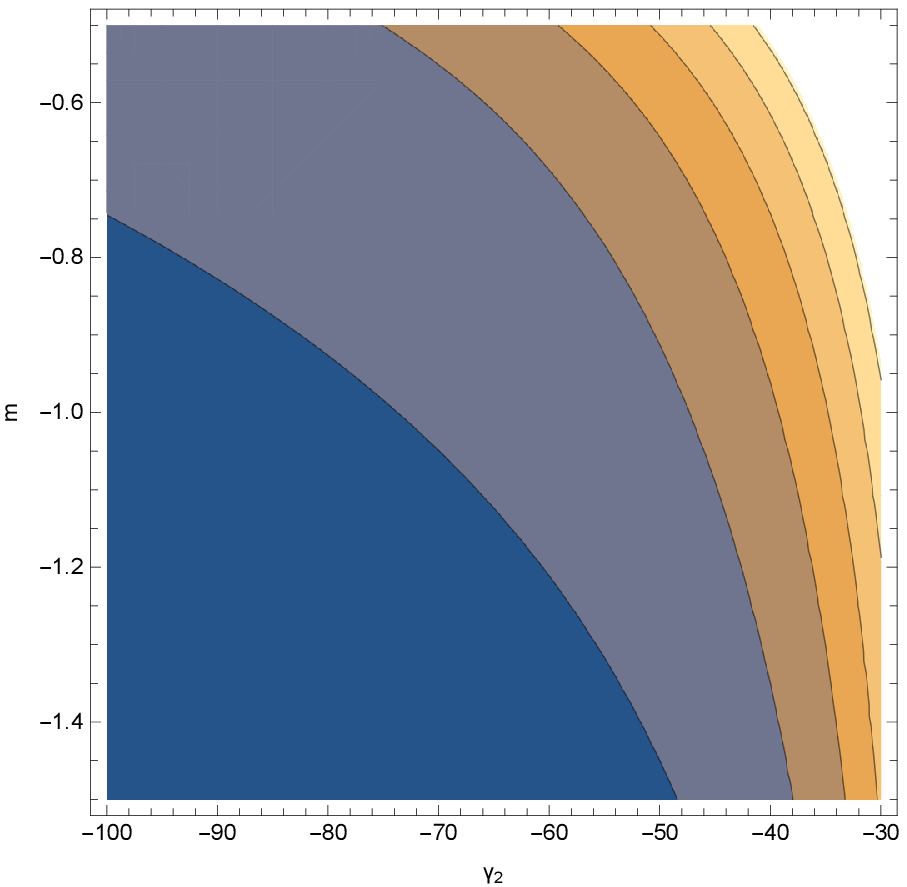}
\includegraphics[width=3.5pc]{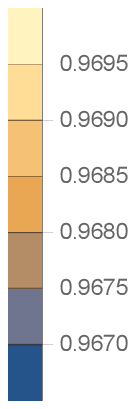}
\includegraphics[width=17pc]{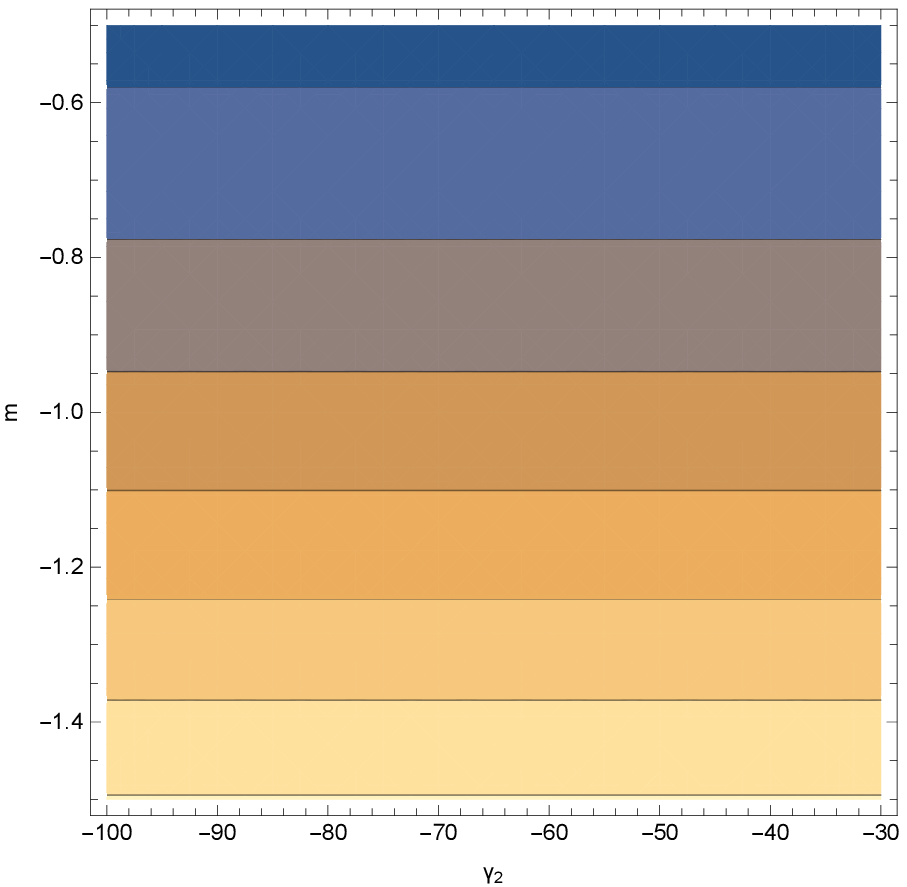}
\includegraphics[width=2.5pc]{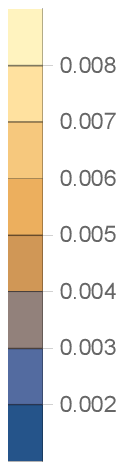}
\caption{Scalar spectral index of primordial curvature
perturbations $n_S$ (left) and tensor-to-scalar ratio $r$ (right)
as functions of $\gamma_2$ and $m$ which range take values
[-100,-35] and [-1.5,-0.5] respectively. These plots correspond to
Eq. (\ref{motion4}) being the one used. In this case, the scalar
spectral index depends on both parameters whereas the
tensor-to-scalar ratio keeps being affected only by the power-law
exponent $m$.}
\end{figure}

Since $V'\sim\mathcal{O}(10)$ while $3\omega
H\dot\phi\sim\mathcal{O}(10^{-3})$, the continuity equation can be
further simplified. Discarding $3\omega H\dot\phi$ from Eq.
(\ref{motion3B}) leads to a power-law scalar potential which is
similar to the previous model. Also, even if the reader might feel
unsatisfied with the numerical value of the slow-roll indices
during the first horizon crossing, we mention that by using Eq.
(\ref{motion4}) as Hubble's parameter results in a perplexed
scalar potential comprised of hypergeometric functions as shown
below.
\begin{equation}
\centering
\label{VB2}
V(\phi)=V_2(4m\gamma_2+(\frac{c}{\kappa}+4\gamma_2^2\kappa)\phi)^{2m}e^{-\frac{4\gamma_2\kappa^2\omega\left(-1+_2F_1\left(1, 1-m, 2-m; -\frac{((2\gamma_2\kappa)^2+c)\phi}{4m\gamma_2\kappa}\right)\right)}{(m-1)\Lambda_2((2\gamma_2\kappa)^2+c)}}\, ,
\end{equation}
In this case, in order to obtain a viable phenomenology, a
redefinition of the free parameters is needed. Reassigning the
values $m=-0.5$ and $\gamma_2=-60$ restores compatibility as now
$n_S=0.967962$, $r=0.00163923$ and $n_T=-0.0002041$ are
acceptable. In this case, the orders of magnitude decrease however
the approximations made are still valid, although not in the same
ratios. Referring to the slow-roll indices, we now have
$\epsilon_i\sim\mathcal{O}(10^{-3})$ for the most part,
specifically $\epsilon_1=-0.0040298$,
$\epsilon_3=\epsilon_5=\epsilon_6=0.00413223$ and only
$\epsilon_4=0.0242458$, meaning of order $\mathcal{O}(10^{-2})$.
The hypergeometric function on the other hand might be bizarre or
unappealing, however since $3\omega H\dot\phi\ll V'$, one can
discard such term from Eq. (\ref{motion3B}). Consequently, the
scalar potential acquires once again a power-law form and the
exact same values of free parameters, with the only difference
being $m=-0.5$ and $\gamma_2=-60$, seems to produce the exact same
observational indices, as expected.

Similar to the previous subsection, by discarding $h''\dot\phi^2$
from Eq. (\ref{motion5}), a reasonable assumption as hinted by the
order of magnitudes of the showcased previously, then for
($\omega$, $\lambda_2$, $\Lambda_2$, $V_2$, $N$, $c$, $\gamma_2$,
$m$)=(1, 1, 100, 1, 60, 0.002, 30, 2), the scalar spectral index
is $n_S=0.966667$, and essentially $r=0=n_T$. In this case as
well, no B-modes are expected. The orders of magnitude remain
exactly the same, something which is expected since as stated
before, $h''\dot\phi^2$ is two orders of magnitude lesser than
$H\dot h$. The same applies to the case of $3H\dot h$
participating in Eq. (\ref{motion4}) and for the scalar potential
comprised of hypergeometric function or not and for the values
($\omega$, $\lambda_2$, $\Lambda_2$, $V_2$, $N$, $c$, $\gamma_2$,
$m$)=(1, 1, 100, 1, 60, 0.002, -60, -0.5), meaning that B-modes
are absent. In both cases for the coupling functions, it seems
that the choice for Eq. (\ref{motion2}) determines and affects the
existence of B-modes. Thus, this feature is a characteristic of
the exponential Gauss-Bonnet coupling, since it is the only
unchanged function from the previous model.

\subsection{Linear Gauss-Bonnet And Power-Law Ricci Coupling}

As a final model, we shall study another simple case for the
coupling functions, which happens to facilitate our study greatly.
Let the Gauss-Bonnet and Ricci scalar non-minimal couplings be
chosen as follows,
\begin{equation}
\centering
\label{xiC}
\xi(\phi)=\lambda_3\kappa\phi\, ,
\end{equation}
\begin{equation}
\centering
\label{hC}
h(\phi)=\Lambda_3(\kappa\phi)^n\, ,
\end{equation}
This is an interesting choice for the coupling functions due to
the fact that the specific choice for $h$ simplifies the ratios
$h'/h$ and $h''/h$ which appear in Eq. (\ref{motion2}) but also
the linear Gauss-Bonnet coupling simplifies $\dot\phi$ from Eq.
(\ref{dotphi}) since $\xi''=0$. The linear Gauss-Bonnet coupling
was proved to be inconsistent with the observations in the case of
having solely the Gauss-Bonnet term $\xi(\phi)\mathcal{G}$ in the
Lagrangian, while on the other hand it provided viable results by
further assuming the constant-roll assumption for the minimally
coupled case $h(\phi)=1$, hence it is interesting to examine this
choice as well. In addition, we shall make use of the following
equations of motion,
\begin{equation}
\centering
\label{motion1C}
H^2=\frac{\kappa^2V}{3h\left(1+\frac{h'}{h}\frac{4\xi'}{4\xi''+c\xi}\right)}\, ,
\end{equation}
\begin{equation}
\centering
\label{motion2C}
\dot H=-\frac{2H^2\xi'}{4\xi''+c\xi}\left(\frac{h''}{h}\frac{4\xi'}{4\xi''+c\xi}-\frac{h'}{h}\right)\, ,
\end{equation}
\begin{equation}
\centering
\label{motion3C}
V'-6H^2\frac{h'}{\kappa^2}=0\, ,
\end{equation}
and in the next section we shall overview all the possible
approximations that can be performed in order to simplify the
equations of motion in the above form.

In the equations of motion, the term $\xi''$ was kept for the sake
of completeness but essentially due to the linear Gauss-Bonnet
coupling its zero. For the scalar potential, since the
aforementioned equations are used, then subsequently we have,
\begin{equation}
\centering
\label{VC}
V(\phi)=V_3(4n+c\phi^2)^n\, ,
\end{equation}
In this particular approach, each and every scalar function has a
power-law dependence on the scalar field, with the exponents being
not necessarily integers , but we shall use only integer values.
In addition, the first three slow-roll indices are written as,
\begin{equation}
\centering
\label{index1C}
\epsilon_1=-2n\frac{c\phi^2-4n+4}{(c\phi^2)^2}\, ,
\end{equation}
\begin{equation}
\centering
\label{index2C}
\epsilon_2=2\frac{c(n-2)\phi^2-4n(n-1)}{(c\phi^2)^2}\, ,
\end{equation}
\begin{equation}
\centering \label{index3C} \epsilon_3=\frac{2n}{c\phi^2}\, .
\end{equation}
From $\epsilon_1$ and Eq. (\ref{efolds}), one easily obtains the
expressions for the scalar field,
\begin{equation}
\centering
\label{phiiC}
\phi_i=\sqrt{\frac{c\phi_f^2-8N}{c}}\, ,
\end{equation}
\begin{equation}
\centering
\label{phifC}
\phi_f=\sqrt{-\frac{n}{c}+\sqrt{\frac{n(9n-8)}{c^2}}}\, ,
\end{equation}
For $\phi_f$, since it is derived from the relation
$\epsilon_1=1$, there exist 4 possible expressions but we shall
limit our work only to this particular. In consequence,
designating ($\omega$, $\lambda_3$, $\Lambda_3$, $V_3$, $N$, $c$,
$n$)=(1, 1, 100, 1, 60, -0.01, 2) then the observational indices
take the values $n_S=0.966395$, $n_T=-0.0001389$ and $r=0.0011023$
which are compatible with the latest Planck 2018 data
\cite{Akrami:2018odb}. The model is free of ghost instabilities,
since $c_A=1$ and finally, $\epsilon_1=0.00829$,
$\epsilon_2=-0.000068$, $\epsilon_3=-0.00822246$,
$\epsilon_4=-0.00822687$, $\epsilon_5=-0.00822249$ and
$\epsilon_6=\epsilon_3$ at horizon crossing. These values are
indicative of the validity of the slow-roll approximations, but it
is worth stating that even though that essentially
$\epsilon_5\simeq\epsilon_6$, the string corrections seem to have
an impact on the dynamics of the model even for trivial values of
the free parameters. It is expected that their contribution is not
as negligible as previously.

Let us now proceed with the validity of the approximations made
throughout the equations of motion. As it was hinted previously,
the slow-roll conditions apply as $\dot H\sim\mathcal{O}(10^{-4})$
while $H^2\sim\mathcal{O}(10^{-2})$,
$\frac{1}{2}\omega\dot\phi^2\sim\mathcal{O}(10^{-2})$ in contrast
to $V\sim\mathcal{O}(10^5)$ and finally,
$\ddot\phi\sim\mathcal{O}(10^{-6})$ whereas
$H\dot\phi\sim\mathcal{O}(10^{-2})$. The string corrections now
seem to be enhanced compared to the previous two examples but
still inferior. Particularly, $24\dot\xi
H^3\sim\mathcal{O}(10^{-2})$,
$9c\xi\dot\phi^2H^2\sim\mathcal{O}(10^{-2})$, $16\dot\xi H\dot
H\sim\mathcal{O}(10^{-5})$,
$8H^2(\ddot\xi-H\dot\xi)\sim\mathcal{O}(10^{-3})$,
$c\dot\phi\left(2\xi\dot\phi(\dot
H-3H^2)+4H\xi\ddot\phi+2H\dot\xi\dot\phi\right)\sim\mathcal{O}(10^{-2})$,
$\xi'\mathcal{G}\sim\mathcal{O}(10^{-3})$ and
$3c\left(H^2(\dot\xi\dot H+2\xi\ddot\phi)+2H(2\dot
H+3H^2)\xi\dot\phi\right)\sim\mathcal{O}(10^{-2})$ which are all
negligible compared to the dominant terms as stated before.
Finally, we have $3H\dot h\sim\mathcal{O}(10^3)$ and
$V\sim\mathcal{O}(10^5)$ thus the first could in principle be
neglected, $\omega\dot\phi^2\sim\mathcal{O}(10^{-2})$,
$h''\dot\phi^2\sim\mathcal{O}(10)$ and $H\dot
h\sim\mathcal{O}(10^3)$, meaning that once again $h''\dot\phi^2$
could be neglected, and finally $V'\sim\mathcal{O}(10^3)$ while
$3\omega H\dot\phi\sim\mathcal{O}(10^{-2})$ thus all the
approximations made are indeed valid.

As a final task, it is worth discussing some important issues.
Firstly, if one were to keep the term $3H\omega\dot\phi$ in the
continuity equation, as it was the case with the previous two
examples, then this would be intrinsically unrealistic, since even
though a different scalar potential would be produced, which does
not influence the results, for the sake of completeness one would
have to add at least the string correction
$3c\left(H^2(\dot\xi\dot H+2\xi\ddot\phi)+2H(2\dot
H+3H^2)\xi\dot\phi\right)$ since it is of the same order, in
particular $\mathcal{O}(10^{-2})$. Moreover, neglecting $3H\dot h$
from Eq. (\ref{motion1C}) and by either increasing by one order
$\Lambda_3$ or decreasing $\lambda_3$, produces a viable
phenomenology. For instance, $\Lambda_3=10^3$ gives
$n_S=0.967562$, $r=0.00111947$ and $n_T=-0.00013637$ which
genuinely speaking are not dramatic changes. The only significant
change is $\dot\phi$ being independent of $\phi$, therefore the
evolution of the scalar field follows a linear law
$\phi(t)=\phi_i+\frac{4}{c}\sqrt{\frac{V_3}{3\Lambda_1}}t$ thus
the tine instance of the final stage of the inflationary era can
be easily extracted. The scalar potential also experiences a mild
change since the term $4n$ in Eq. (\ref{VC}) vanishes. In contrast
to the previous cases, neglecting $h''\dot\phi^2$ from Eq.
(\ref{motion2C}) does not imply that B-modes are absent. Indeed,
both the tensor-to-scalar ratio and the tensor spectral index are
quite small as $r=3.8\cdot10^{-7}$ and $n_T=-1.9\cdot10^{-10}$ but
for instance decreasing $\Lambda_3$ to $\Lambda_3=6$ leads to
$n_S=0.968855$ and $r=0.000105$ which are obviously acceptable
values. The same applies to the case of neglecting both
$h''\dot\phi^2$ and $3H\dot h$ from equations (\ref{motion2C}) and
(\ref{motion1C}) respectively.

In consequence, the linear Gauss-Bonnet coupling can work either
in the minimally or non-minimally coupled case, under the
slow-roll and constant-roll assumption only if extra string
corrections are assumed to be present.

\section{A Panorama of Different Assumptions and Simplifications that Can be Performed for the Sake of Analyticity}

In the previous sections, it was stated that there exist several
possible forms that the gravitational equations of motion can
take, under different assumptions. In principle none of them is
intrinsically unrealistic from the beginning, unless a specific
model is incapable of producing phenomenologies compatible with
the observational data, after the free parameters of the model are
specified. In all the models studied so far, we showcased only
some of the possible configurations of the equations of motion due
to the fact that there exist multiple paths one can follow for a
single model function. Thus, it is worth devoting a section where
we present the possible approximated forms of the equations of
motion along with the relations that must be satisfied in order
for a model to be rendered successful phenomenologically. Before
we proceed however, it is worth relabelling the string correction
terms in order not to obtain lengthy forms. In particular, we
denote string corrections as,
\begin{equation}
\centering
\label{S1}
S_1(\phi)=24\dot\xi H^3-9c\xi H^2\dot\phi^2\, ,
\end{equation}
\begin{equation}
\centering
\label{S2}
S_2=-8H^2(\ddot\xi-H\dot\xi)-16\dot\xi H\dot H+2c\dot\phi\left(\xi(\dot H-3H^2)\dot\phi+2H\xi\ddot\phi+H\dot\xi\dot\phi\right)\, ,
\end{equation}
\begin{equation}
\centering
\label{S3}
S_3=\xi'\mathcal{G}-3c\left(H^2(\dot\xi\dot H+2\xi\ddot\phi)+2H(2\dot H+3H^2)\xi\dot\phi\right)\, ,
\end{equation}
for convenience. Therefore, the main relations that must be
satisfied are the slow-roll conditions,
\begin{align}
\centering
\label{slowrollapprox}
\dot H&\ll H^2&\frac{1}{2}\omega\dot\phi^2&\ll V&\ddot\phi&\ll H\dot\phi\, ,
\end{align}
and the string corrections being inferior compared to the rest
terms, meaning,
\begin{align}
\centering \label{Sapprox} S_1&\ll V&\kappa^2S_2&\ll 2h\dot
H&S_3&\ll V'\, .
\end{align}
The approximations above refer to the order of magnitude and not
the sign obviously and are the core assumptions, or in other words
the same assumptions as in the minimally coupled case. The
equations of motion, even after discarding string corrections
$S_i$ and implementing the slow-roll conditions still contain
extra terms and can be further simplified. Genuinely speaking,
further assumptions are not mandatory but are quite useful if
these are appropriately justified. Let us showcase the necessary
relations in case extra assumptions are assumed in each of the
equations of motion separately.

We commence from the first equation of motion or in other words
Hubble's form. The possible forms are,
\begin{equation}
\centering
\label{H1}
H^2=\frac{\kappa^2V}{3h\left(1+\frac{4\xi'}{4\xi''+c\xi}\frac{h'}{h}\right)}\, ,
\end{equation}
\begin{equation}
\centering
\label{H2}
H^2=\frac{\kappa^2V}{3h}\, ,
\end{equation}
\begin{equation}
\centering \label{H3}
H^2=\frac{\kappa^2V}{3h'}\frac{4\xi''+c\xi}{4\xi'}\, .
\end{equation}
The first two are the ones used mainly in this paper but we
mention also the third choice since one can solve easily for
squared Hubble's parameter. Essentially, the first possible form
of Eq. (\ref{H1}) has no extra assumptions while the other two
have opposite relations which govern the appropriate form. In
particular, we have,
\begin{align}
\centering
\dot h&\ll Hh&Hh&\ll\dot h\, ,
\end{align}
but since the slow-roll conditions are assumed to hold true,
meaning that $\epsilon_3\ll1$ then Eq. (\ref{H3}) is naturally
discarded. Thus, there exist only two possible forms, namely
equations (\ref{H1}) and (\ref{H2}).

Referring to the scalar field equation, there exist once again two
possible forms which one can use. In particular,
\begin{equation}
\centering
\label{V'1}
V'+3H^2\left(\frac{4\omega\xi'}{4\xi''+c\xi}-2\frac{h'}{\kappa^2}\right)=0\, ,
\end{equation}
\begin{equation}
\centering \label{V'2} V'-6H^2\frac{h'}{\kappa^2}=0\, .
\end{equation}
The first form is the one used mainly in this framework while the
latter was used only in the final example. Genuinely speaking, one
is free to also neglect the term $6H^2\frac{h'}{\kappa^2}$ and
thus work with.
\begin{equation}
\centering
\label{V'3}
V'+\omega H^2\frac{12\xi'}{4\xi''+c\xi}=0\, ,
\end{equation}
however this case is used for the non-minimally coupled scenario
so it was not implemented  in the present paper, although it can
be used for a weak Ricci coupling without losing integrity of the
theory, since in either case the Ricci coupling still participates
in Eq. (\ref{motion2}). Therefore, in order for the continuity
equation to be valid, the following expressions must be valid,
\begin{align}
\centering
\kappa^2\dot\phi&\ll 2H h'&2Hh'&\ll\kappa^2\dot\phi\, ,
\end{align}
for Eq. (\ref{V'2}) and Eq. (\ref{V'3}) respectively. As expected,
the conditions are opposite. Finally, for Hubble's derivative and
in consequence the first slow-roll index, there exist $3!$
possible simplified expressions since 3 terms are present in Eq.
(\ref{motion5}). Specifically, we have,
\begin{equation}
\centering
\label{H'1}
\dot H=\frac{2H^2\xi'}{4\xi''+c\xi}\frac{h'}{h}\, ,
\end{equation}
\begin{equation}
\centering
\label{H'2}
\dot H=-\frac{H^2}{2}\left(\frac{4\xi'}{4\xi''+c\xi}\right)^2\frac{h''}{h}\, ,
\end{equation}
\begin{equation}
\centering
\label{H'3}
\dot H=-\frac{\omega H^2}{2h}\left(\frac{4\kappa\xi'}{4\xi''+c\xi}\right)^2\, ,
\end{equation}
\begin{equation}
\centering
\label{H'4}
\dot H=\frac{2H^2\xi'}{4\xi''+c\xi}\left(\frac{h'}{h}-\frac{h''}{h}\frac{4\xi'}{4\xi''+c\xi}\right)\, ,
\end{equation}
\begin{equation}
\centering
\label{H'5}
\dot H=\frac{2H^2\xi'}{4\xi''+c\xi}\left(\frac{h'}{h}-\frac{\kappa^2\omega}{h}\frac{4\xi'}{4\xi''+c\xi}\right)\, ,
\end{equation}
\begin{equation}
\centering \label{H'6} \dot
H=-\frac{H^2}{2}\left(\frac{4\xi'}{4\xi''+c\xi}\right)^2\left(\frac{\kappa^2\omega}{h}+\frac{h''}{h}\right)\,
.
\end{equation}
Essentially, Eq. (\ref{H'3}) is reminiscing of the form of the
minimally coupled case, something that can easily be inferred by
replacing $h(\phi)=1$ and this is the reason that we did not study
this case in the present paper. On the contrary, it was deemed
suitable to examine the contribution of terms proportional to
$\frac{h'}{h}$ and $\frac{h''}{h}$. Once again, it is worth
mentioning that these are all the possible simplified form of Eq.
(\ref{motion5}) since the latter is quite difficult to tackle for
a given $\xi(\phi)$ and $h(\phi)$. Hence, the necessary relations
that must be satisfied for each separate equation in order for a
model to be rendered as viable are,
\begin{align}
\centering
h''\frac{4\xi'}{4\xi''+c\xi}&\ll h'&\kappa^2\omega\frac{4\xi'}{4\xi''+c\xi}&\ll h'\, ,
\end{align}
\begin{align}
\centering
\kappa^2\omega&\ll h''&h'&\ll h''\frac{4\xi'}{4\xi''+c\xi}\, ,
\end{align}
\begin{align}
\centering
h''&\ll\kappa^2\omega&h'&\ll\kappa^2\omega\frac{4\xi'}{4\xi''+c\xi}\, ,
\end{align}
\begin{align}
\centering
\kappa^2\omega&\ll h''&\kappa^2\omega\frac{4\xi'}{4\xi''+c\xi}&\ll h'\, ,
\end{align}
\begin{align}
\centering
h''&\ll\kappa^2\omega&h''\frac{4\xi'}{4\xi''+c\xi}&\ll h'\, ,
\end{align}
\begin{align}
\centering
h'&\ll\kappa^2\omega\frac{4\xi'}{4\xi''+c\xi}&h'&\ll h''\frac{4\xi'}{4\xi''+c\xi}\, ,
\end{align}
for each  $\dot H$ respectively. Due to the slow-roll conditions,
having $h'\ll h''\frac{4\xi'}{4\xi''+c\xi}$ is impossible but such
possibility is still presented for the sake of completeness.
Therefore, there exist essentially 4 viable simplified expressions
for $\dot H$. One must be wary of these relations since they
produce different phenomenologies, as it was demonstrated
previously. Each and every respective relation here must be
satisfied and most importantly, the slow-roll conditions along
with string corrections being inferior to $H^2$ and potential
terms, in order to obtain a viable phenomenology. Even a single
violation results in a false positive model, provided that
compatible results where produced. This result is quite similar to
the case of pure Einstein-Gauss-Bonnet gravity.

\section{Conclusions}

In this work we developed a theoretical framework that practically
revived non-minimally coupled Horndeski theories with non-minimal
kinetic coupling. Particularly we studied kinetic coupling
corrected non-minimally coupled Einstein-Gauss-Bonnet theory, and
we demonstrated how it is possible to have primordial
gravitational waves with propagation speed equal to that of
light's. The effective Lagrangian of this model consists of three
parts, of a canonical scalar field theory non-minimally coupled to
gravity (the Ricci scalar), a non-minimal coupling of the scalar
field to the Gauss-Bonnet invariant, and a non-minimal kinetic
coupling term. This theory without the Gauss-Bonnet invariant term
is excluded by the GW170817 event, but the presence of the
Gauss-Bonnet plays a catalytic role in rendering the whole
theoretical framework compatible with the GW170817 event. The main
constraint we imposed in the theory is the constraint that the
propagation speed of the gravitational tensor perturbations is
equal to unity in natural units. By also exploiting several
slow-roll motivated approximations, we were able to write the
initially quite involved equations of motion, in a simple form.
Eventually we derived the slow-roll indices of the theory, and the
corresponding observations indices. Accordingly, by using several
appropriate models, we explicitly demonstrated how the theoretical
framework of kinetic coupling corrected non-minimally coupled
Einstein-Gauss-Bonnet theory can generate a viable inflationary
era, compatible with the Planck 2018 observational data. For each
model we showed that the compatibility can be achieved for a wide
range of the free parameters of the model. Furthermore, we showed
that all the appropriate approximations and assumptions which we
made in order to derive the equations of motion in a more
analytically tractable form, indeed hold true for all the studied
cases, and for the values of the free parameters that guaranteed
the viability of each model. The next step in all the classes of
the new theoretical framework we developed, is applying the theory
in the context of theoretical astrophysics, and specifically
aiming for the physics of static neutron stars, basically a direct
generalization of \cite{Maselli:2016gxk}. In the literature, both
the scalar Gauss-Bonnet coupling and the scalar field potential
are freely chosen, so it is worth investigating the structure and
evolution of neutron stars for this class of Horndeski theories we
developed, under the constraint that the scalar potential and the
non-minimal Gauss-Bonnet coupling are constrained. Also, the
presence of the non-minimal kinetic coupling may eventually play
an important role, so we aim in the near future to adopt the line
of research we just described.


\begin{thebibliography}{99}














\bibitem{TheLIGOScientific:2017qsa}
B.~P.~Abbott \textit{et al.} [LIGO Scientific and Virgo],
Phys. Rev. Lett. \textbf{119} (2017) no.16, 161101
doi:10.1103/PhysRevLett.119.161101 [arXiv:1710.05832 [gr-qc]].



\bibitem{GBM:2017lvd}
  B.~P.~Abbott {\it et al.}
  ``Multi-messenger Observations of a Binary Neutron Star Merger,''
  Astrophys.\ J.\  {\bf 848} (2017) no.2,  L12
  doi:10.3847/2041-8213/aa91c9
  [arXiv:1710.05833 [astro-ph.HE]].




\bibitem{Ezquiaga:2017ekz}
J.~M.~Ezquiaga and M.~Zumalac\'arregui,
Phys. Rev. Lett. \textbf{119} (2017) no.25, 251304
doi:10.1103/PhysRevLett.119.251304 [arXiv:1710.05901
[astro-ph.CO]].


\bibitem{Baker:2017hug}
T.~Baker, E.~Bellini, P.~G.~Ferreira, M.~Lagos, J.~Noller and
I.~Sawicki,
Phys. Rev. Lett. \textbf{119} (2017) no.25, 251301
doi:10.1103/PhysRevLett.119.251301 [arXiv:1710.06394
[astro-ph.CO]].


\bibitem{Creminelli:2017sry}
P.~Creminelli and F.~Vernizzi,
Phys. Rev. Lett. \textbf{119} (2017) no.25, 251302
doi:10.1103/PhysRevLett.119.251302 [arXiv:1710.05877
[astro-ph.CO]].


\bibitem{Sakstein:2017xjx}
J.~Sakstein and B.~Jain,
Phys. Rev. Lett. \textbf{119} (2017) no.25, 251303
doi:10.1103/PhysRevLett.119.251303 [arXiv:1710.05893
[astro-ph.CO]].





\bibitem{horndeskioriginal} G. W. Horndeski, Int.J.Theor.Phys. 10, 363 (1974).



\bibitem{Kobayashi:2019hrl}
T.~Kobayashi,
Rept. Prog. Phys. \textbf{82} (2019) no.8, 086901
doi:10.1088/1361-6633/ab2429 [arXiv:1901.07183 [gr-qc]].



\bibitem{Kobayashi:2016xpl}
T.~Kobayashi,
Phys. Rev. D \textbf{94} (2016) no.4, 043511
doi:10.1103/PhysRevD.94.043511 [arXiv:1606.05831 [hep-th]].



\bibitem{Crisostomi:2016tcp}
M.~Crisostomi, M.~Hull, K.~Koyama and G.~Tasinato,
JCAP \textbf{03} (2016), 038 doi:10.1088/1475-7516/2016/03/038
[arXiv:1601.04658 [hep-th]].


\bibitem{Bellini:2015xja}
E.~Bellini, A.~J.~Cuesta, R.~Jimenez and L.~Verde,
JCAP \textbf{02} (2016), 053 doi:10.1088/1475-7516/2016/06/E01
[arXiv:1509.07816 [astro-ph.CO]].

\bibitem{Gleyzes:2014qga}
J.~Gleyzes, D.~Langlois, F.~Piazza and F.~Vernizzi,
JCAP \textbf{02} (2015), 018 doi:10.1088/1475-7516/2015/02/018
[arXiv:1408.1952 [astro-ph.CO]].




\bibitem{Lin:2014jga}
C.~Lin, S.~Mukohyama, R.~Namba and R.~Saitou,
JCAP \textbf{10} (2014), 071 doi:10.1088/1475-7516/2014/10/071
[arXiv:1408.0670 [hep-th]].






\bibitem{Deffayet:2013lga}
C.~Deffayet and D.~A.~Steer,
Class. Quant. Grav. \textbf{30} (2013), 214006
doi:10.1088/0264-9381/30/21/214006 [arXiv:1307.2450 [hep-th]].



\bibitem{Bettoni:2013diz}
D.~Bettoni and S.~Liberati,
Phys. Rev. D \textbf{88} (2013), 084020
doi:10.1103/PhysRevD.88.084020 [arXiv:1306.6724 [gr-qc]].



\bibitem{Koyama:2013paa}
K.~Koyama, G.~Niz and G.~Tasinato,
Phys. Rev. D \textbf{88} (2013), 021502
doi:10.1103/PhysRevD.88.021502 [arXiv:1305.0279 [hep-th]].





\bibitem{Starobinsky:2016kua}
A.~A.~Starobinsky, S.~V.~Sushkov and M.~S.~Volkov,
JCAP \textbf{06} (2016), 007 doi:10.1088/1475-7516/2016/06/007
[arXiv:1604.06085 [hep-th]].



\bibitem{Capozziello:2018gms}
S.~Capozziello, K.~F.~Dialektopoulos and S.~V.~Sushkov,
Eur. Phys. J. C \textbf{78} (2018) no.6, 447
doi:10.1140/epjc/s10052-018-5939-1 [arXiv:1803.01429 [gr-qc]].


\bibitem{BenAchour:2016fzp}
J.~Ben Achour, M.~Crisostomi, K.~Koyama, D.~Langlois, K.~Noui and
G.~Tasinato,
JHEP \textbf{12} (2016), 100 doi:10.1007/JHEP12(2016)100
[arXiv:1608.08135 [hep-th]].



\bibitem{Starobinsky:2019xdp}
A.~A.~Starobinsky, S.~V.~Sushkov and M.~S.~Volkov,
Phys. Rev. D \textbf{101} (2020) no.6, 064039
doi:10.1103/PhysRevD.101.064039 [arXiv:1912.12320 [hep-th]].













\bibitem{Sushkov:2009hk}
S.~V.~Sushkov,
Phys. Rev. D \textbf{80} (2009), 103505
doi:10.1103/PhysRevD.80.103505 [arXiv:0910.0980 [gr-qc]].


\bibitem{Minamitsuji:2013ura}
M.~Minamitsuji,
Phys. Rev. D \textbf{89} (2014), 064017
doi:10.1103/PhysRevD.89.064017 [arXiv:1312.3759 [gr-qc]].



\bibitem{Saridakis:2010mf}
E.~N.~Saridakis and S.~V.~Sushkov,
Phys. Rev. D \textbf{81} (2010), 083510
doi:10.1103/PhysRevD.81.083510 [arXiv:1002.3478 [gr-qc]].



\bibitem{Barreira:2013jma}
A.~Barreira, B.~Li, A.~Sanchez, C.~M.~Baugh and S.~Pascoli,
Phys. Rev. D \textbf{87} (2013), 103511
doi:10.1103/PhysRevD.87.103511 [arXiv:1302.6241 [astro-ph.CO]].


\bibitem{Sushkov:2012za}
S.~Sushkov,
Phys. Rev. D \textbf{85} (2012), 123520
doi:10.1103/PhysRevD.85.123520 [arXiv:1204.6372 [gr-qc]].




\bibitem{Barreira:2012kk}
A.~Barreira, B.~Li, C.~M.~Baugh and S.~Pascoli,
Phys. Rev. D \textbf{86} (2012), 124016
doi:10.1103/PhysRevD.86.124016 [arXiv:1208.0600 [astro-ph.CO]].


\bibitem{Skugoreva:2013ooa}
M.~A.~Skugoreva, S.~V.~Sushkov and A.~V.~Toporensky,
Phys. Rev. D \textbf{88} (2013), 083539
doi:10.1103/PhysRevD.88.083539 [arXiv:1306.5090 [gr-qc]].


\bibitem{Gubitosi:2011sg}
G.~Gubitosi and E.~V.~Linder,
Phys. Lett. B \textbf{703} (2011), 113-118
doi:10.1016/j.physletb.2011.07.066 [arXiv:1106.2815
[astro-ph.CO]].






\bibitem{Matsumoto:2015hua}
J.~Matsumoto and S.~V.~Sushkov,
JCAP \textbf{11} (2015), 047 doi:10.1088/1475-7516/2015/11/047
[arXiv:1510.03264 [gr-qc]].

\bibitem{Deffayet:2010qz}
C.~Deffayet, O.~Pujolas, I.~Sawicki and A.~Vikman,
JCAP \textbf{10} (2010), 026 doi:10.1088/1475-7516/2010/10/026
[arXiv:1008.0048 [hep-th]].



\bibitem{Granda:2010hb}
L.~Granda and W.~Cardona,
JCAP \textbf{07} (2010), 021 doi:10.1088/1475-7516/2010/07/021
[arXiv:1005.2716 [hep-th]].



\bibitem{Matsumoto:2017gnx}
J.~Matsumoto and S.~V.~Sushkov,
JCAP \textbf{01} (2018), 040 doi:10.1088/1475-7516/2018/01/040
[arXiv:1703.04966 [gr-qc]].



\bibitem{Gao:2010vr}
C.~Gao,
JCAP \textbf{06} (2010), 023 doi:10.1088/1475-7516/2010/06/023
[arXiv:1002.4035 [gr-qc]].


\bibitem{Granda:2009fh}
L.~Granda,
JCAP \textbf{07} (2010), 006 doi:10.1088/1475-7516/2010/07/006
[arXiv:0911.3702 [hep-th]].

\bibitem{Germani:2010gm}
C.~Germani and A.~Kehagias,
Phys. Rev. Lett. \textbf{105} (2010), 011302
doi:10.1103/PhysRevLett.105.011302 [arXiv:1003.2635 [hep-ph]].


\bibitem{Fu:2019ttf}
C.~Fu, P.~Wu and H.~Yu,
Phys. Rev. D \textbf{100} (2019) no.6, 063532
doi:10.1103/PhysRevD.100.063532 [arXiv:1907.05042 [astro-ph.CO]].






\bibitem{Hwang:2005hb}
  J.~c.~Hwang and H.~Noh,
  Phys.\ Rev.\ D {\bf 71} (2005) 063536
  doi:10.1103/PhysRevD.71.063536
  [gr-qc/0412126].


\bibitem{Nojiri:2006je}
  S.~Nojiri, S.~D.~Odintsov and M.~Sami,
  Phys.\ Rev.\ D {\bf 74} (2006) 046004
  doi:10.1103/PhysRevD.74.046004
  [hep-th/0605039].




\bibitem{Cognola:2006sp}
  G.~Cognola, E.~Elizalde, S.~Nojiri, S.~Odintsov and S.~Zerbini,
  Phys.\ Rev.\ D {\bf 75} (2007) 086002
  doi:10.1103/PhysRevD.75.086002
  [hep-th/0611198].



\bibitem{Nojiri:2005vv}
  S.~Nojiri, S.~D.~Odintsov and M.~Sasaki,
  Phys.\ Rev.\ D {\bf 71} (2005) 123509
  doi:10.1103/PhysRevD.71.123509
  [hep-th/0504052].


\bibitem{Nojiri:2005jg}
  S.~Nojiri and S.~D.~Odintsov,
  Phys.\ Lett.\ B {\bf 631} (2005) 1
  doi:10.1016/j.physletb.2005.10.010
  [hep-th/0508049].







\bibitem{Satoh:2007gn}
  M.~Satoh, S.~Kanno and J.~Soda,
  Phys.\ Rev.\ D {\bf 77} (2008) 023526
  doi:10.1103/PhysRevD.77.023526
  [arXiv:0706.3585 [astro-ph]].



\bibitem{Bamba:2014zoa}
  K.~Bamba, A.~N.~Makarenko, A.~N.~Myagky and S.~D.~Odintsov,
  JCAP {\bf 1504} (2015) 001
  doi:10.1088/1475-7516/2015/04/001
  [arXiv:1411.3852 [hep-th]].


\bibitem{Yi:2018gse}
  Z.~Yi, Y.~Gong and M.~Sabir,
  Phys.\ Rev.\ D {\bf 98} (2018) no.8,  083521
  doi:10.1103/PhysRevD.98.083521
  [arXiv:1804.09116 [gr-qc]].


\bibitem{Guo:2009uk}
  Z.~K.~Guo and D.~J.~Schwarz,
  Phys.\ Rev.\ D {\bf 80} (2009) 063523
  doi:10.1103/PhysRevD.80.063523
  [arXiv:0907.0427 [hep-th]].


\bibitem{Guo:2010jr}
  Z.~K.~Guo and D.~J.~Schwarz,
  Phys.\ Rev.\ D {\bf 81} (2010) 123520
  doi:10.1103/PhysRevD.81.123520
  [arXiv:1001.1897 [hep-th]].


\bibitem{Jiang:2013gza}
  P.~X.~Jiang, J.~W.~Hu and Z.~K.~Guo,
  Phys.\ Rev.\ D {\bf 88} (2013) 123508
  doi:10.1103/PhysRevD.88.123508
  [arXiv:1310.5579 [hep-th]].



\bibitem{Kanti:2015pda}
  P.~Kanti, R.~Gannouji and N.~Dadhich,
  Phys.\ Rev.\ D {\bf 92} (2015) no.4,  041302
  doi:10.1103/PhysRevD.92.041302
  [arXiv:1503.01579 [hep-th]].


\bibitem{vandeBruck:2017voa}
  C.~van de Bruck, K.~Dimopoulos, C.~Longden and C.~Owen,
  arXiv:1707.06839 [astro-ph.CO].



\bibitem{Kanti:1998jd}
  P.~Kanti, J.~Rizos and K.~Tamvakis,
  Phys.\ Rev.\ D {\bf 59} (1999) 083512
  doi:10.1103/PhysRevD.59.083512
  [gr-qc/9806085].




\bibitem{Pozdeeva:2020apf}
  E.~O.~Pozdeeva, M.~R.~Gangopadhyay, M.~Sami, A.~V.~Toporensky and S.~Y.~Vernov,
  arXiv:2006.08027 [gr-qc].

\bibitem{Fomin:2020hfh}
  I.~Fomin,
  arXiv:2004.08065 [gr-qc].

\bibitem{DeLaurentis:2015fea}
  M.~De Laurentis, M.~Paolella and S.~Capozziello,
  Phys.\ Rev.\ D {\bf 91} (2015) no.8,  083531
  doi:10.1103/PhysRevD.91.083531
  [arXiv:1503.04659 [gr-qc]].


\bibitem{Chervon:2019sey}
  S.~Chervon, I.~Fomin, V.~Yurov and A.~Yurov,
  doi:10.1142/11405



\bibitem{Nozari:2017rta}
  K.~Nozari and N.~Rashidi,
  Phys.\ Rev.\ D {\bf 95} (2017) no.12,  123518
  doi:10.1103/PhysRevD.95.123518
  [arXiv:1705.02617 [astro-ph.CO]].




\bibitem{Odintsov:2018zhw}
  S.~D.~Odintsov and V.~K.~Oikonomou,
  Phys.\ Rev.\ D {\bf 98} (2018) no.4,  044039
  doi:10.1103/PhysRevD.98.044039
  [arXiv:1808.05045 [gr-qc]].


  \bibitem{Kawai:1998ab}
  S.~Kawai, M.~a.~Sakagami and J.~Soda,
  Phys.\ Lett.\ B {\bf 437}, 284 (1998)
  doi:10.1016/S0370-2693(98)00925-3
  [gr-qc/9802033].


\bibitem{Yi:2018dhl}
  Z.~Yi and Y.~Gong,
  Universe {\bf 5} (2019) no.9,  200
  doi:10.3390/universe5090200
  [arXiv:1811.01625 [gr-qc]].


\bibitem{vandeBruck:2016xvt}
  C.~van de Bruck, K.~Dimopoulos and C.~Longden,
  Phys.\ Rev.\ D {\bf 94} (2016) no.2,  023506
  doi:10.1103/PhysRevD.94.023506
  [arXiv:1605.06350 [astro-ph.CO]].


\bibitem{Kleihaus:2019rbg}
  B.~Kleihaus, J.~Kunz and P.~Kanti,
  arXiv:1910.02121 [gr-qc].





\bibitem{Bakopoulos:2019tvc}
  A.~Bakopoulos, P.~Kanti and N.~Pappas,
  Phys.\ Rev.\ D {\bf 101} (2020) no.4,  044026
  doi:10.1103/PhysRevD.101.044026
  [arXiv:1910.14637 [hep-th]].


\bibitem{Maeda:2011zn}
  K.~i.~Maeda, N.~Ohta and R.~Wakebe,
  Eur.\ Phys.\ J.\ C {\bf 72} (2012) 1949
  doi:10.1140/epjc/s10052-012-1949-6
  [arXiv:1111.3251 [hep-th]].






\bibitem{Bakopoulos:2020dfg}
  A.~Bakopoulos, P.~Kanti and N.~Pappas,
  arXiv:2003.02473 [hep-th].


\bibitem{Ai:2020peo}
W.~Ai,
[arXiv:2004.02858 [gr-qc]].



\bibitem{Odintsov:2019clh}
  S.~D.~Odintsov and V.~K.~Oikonomou,
  Phys.\ Lett.\ B {\bf 797} (2019) 134874
  doi:10.1016/j.physletb.2019.134874
  [arXiv:1908.07555 [gr-qc]].



\bibitem{Oikonomou:2020oil}
V.~K.~Oikonomou and F.~P.~Fronimos,
[arXiv:2007.11915 [gr-qc]].

\bibitem{Odintsov:2020xji}
S.~D.~Odintsov, V.~K.~Oikonomou and F.~P.~Fronimos,
Annals Phys. \textbf{420} (2020), 168250
doi:10.1016/j.aop.2020.168250 [arXiv:2007.02309 [gr-qc]].



\bibitem{Oikonomou:2020sij}
V.~K.~Oikonomou and F.~P.~Fronimos,
[arXiv:2006.05512 [gr-qc]].



\bibitem{Odintsov:2020zkl}
S.~D.~Odintsov and V.~K.~Oikonomou,
Phys. Lett. B \textbf{805} (2020), 135437
doi:10.1016/j.physletb.2020.135437 [arXiv:2004.00479 [gr-qc]].


\bibitem{Odintsov:2020sqy}
S.~D.~Odintsov, V.~K.~Oikonomou and F.~P.~Fronimos,
[arXiv:2003.13724 [gr-qc]].




\bibitem{Odintsov:2020mkz}
S.~D.~Odintsov, V.~K.~Oikonomou, F.~P.~Fronimos and
S.~A.~Venikoudis,
Phys. Dark Univ. \textbf{30} (2020), 100718
doi:10.1016/j.dark.2020.100718 [arXiv:2009.06113 [gr-qc]].


\bibitem{Easther:1996yd}
  R.~Easther and K.~i.~Maeda,
  Phys.\ Rev.\ D {\bf 54} (1996) 7252
  doi:10.1103/PhysRevD.54.7252
  [hep-th/9605173].

\bibitem{Antoniadis:1993jc}
  I.~Antoniadis, J.~Rizos and K.~Tamvakis,
  Nucl.\ Phys.\ B {\bf 415} (1994) 497
  doi:10.1016/0550-3213(94)90120-1
  [hep-th/9305025].

\bibitem{Antoniadis:1990uu}
I.~Antoniadis, C.~Bachas, J.~R.~Ellis and D.~V.~Nanopoulos,
Phys.\ Lett.\ B \textbf{257} (1991), 278-284
doi:10.1016/0370-2693(91)91893-Z




\bibitem{Kanti:1995vq}
P.~Kanti, N.~Mavromatos, J.~Rizos, K.~Tamvakis and E.~Winstanley,
Phys. Rev. D \textbf{54} (1996), 5049-5058
doi:10.1103/PhysRevD.54.5049 [arXiv:hep-th/9511071 [hep-th]].



\bibitem{Kanti:1997br}
P.~Kanti, N.~Mavromatos, J.~Rizos, K.~Tamvakis and E.~Winstanley,
Phys. Rev. D \textbf{57} (1998), 6255-6264
doi:10.1103/PhysRevD.57.6255 [arXiv:hep-th/9703192 [hep-th]].




\bibitem{Akrami:2018odb}
  Y.~Akrami {\it et al.} [Planck Collaboration],
  arXiv:1807.06211 [astro-ph.CO].









\bibitem{Maselli:2016gxk}
A.~Maselli, H.~O.~Silva, M.~Minamitsuji and E.~Berti,
Phys. Rev. D \textbf{93} (2016) no.12, 124056
doi:10.1103/PhysRevD.93.124056 [arXiv:1603.04876 [gr-qc]].















\end{thebibliography}
\end{document}